\documentclass[twocolumn,traditabstract]{aa}

\usepackage{graphicx}
\usepackage{float} 
\usepackage{subfigure}
\usepackage{amsmath}
\usepackage{amsfonts}
\usepackage{amssymb}
\usepackage{txfonts}
\usepackage{natbib}
\usepackage{color}
\definecolor{linkcolor}{rgb}{0.6,0,0}
\definecolor{citecolor}{rgb}{0,0,0.75}
\definecolor{urlcolor}{rgb}{0.12,0.46,0.7}
\usepackage[breaklinks, colorlinks, urlcolor=urlcolor,
    linkcolor=linkcolor,citecolor=citecolor,pdfencoding=auto]{hyperref}
\hypersetup{linktocpage}

\usepackage{enumitem}

\newcommand{\WMAP}{\emph{WMAP}}

\newcommand{\Planck}{\emph{Planck}}

\newcommand{\name}{GreenPol}

\def\setsymbol#1#2{\expandafter\def\csname #1\endcsname{#2}}
\def\getsymbol#1{\csname #1\endcsname}

\def\Planck{\textit{Planck}}

\newbox\tablebox    \newdimen\tablewidth
\def\leaderfil{\leaders\hbox to 5pt{\hss.\hss}\hfil}
\def\endPlancktablewide{\tablewidth=\textwidth 
    $$\hss\copy\tablebox\hss$$
    \vskip-\lastskip\vskip -2pt}
\def\tablenote#1 #2\par{\begingroup \parindent=0.8em
    \abovedisplayshortskip=0pt\belowdisplayshortskip=0pt
    \noindent
    $$\hss\vbox{\hsize\tablewidth \hangindent=\parindent \hangafter=1 \noindent
    \hbox to \parindent{$^#1$\hss}\strut#2\strut\par}\hss$$
    \endgroup}
\def\doubleline{\vskip 3pt\hrule \vskip 1.5pt \hrule \vskip 5pt}

\def\L2{\ifmmode L_2\else $L_2$\fi}

\def\DeltaT{\ifmmode \Delta T\else $\Delta T$\fi}
\def\deltat{\ifmmode \Delta t\else $\Delta t$\fi}
\def\fknee{\ifmmode f_{\rm knee}\else $f_{\rm knee}$\fi}
\def\Fmax{\ifmmode F_{\rm max}\else $F_{\rm max}$\fi}
\def\solar{\ifmmode{\rm M}_{\mathord\odot}\else${\rm M}_{\mathord\odot}$\fi}
\def\Msolar{\ifmmode{\rm M}_{\mathord\odot}\else${\rm M}_{\mathord\odot}$\fi}
\def\Lsolar{\ifmmode{\rm L}_{\mathord\odot}\else${\rm L}_{\mathord\odot}$\fi}
\def\inv{\ifmmode^{-1}\else$^{-1}$\fi}
\def\mo{\ifmmode^{-1}\else$^{-1}$\fi}
\def\sup#1{\ifmmode ^{\rm #1}\else $^{\rm #1}$\fi}
\def\expo#1{\ifmmode \times 10^{#1}\else $\times 10^{#1}$\fi}
\def\,{\thinspace}
\def\lsim{\mathrel{\raise .4ex\hbox{\rlap{$<$}\lower 1.2ex\hbox{$\sim$}}}}
\def\gsim{\mathrel{\raise .4ex\hbox{\rlap{$>$}\lower 1.2ex\hbox{$\sim$}}}}

\def\simprop{\mathrel{\raise .4ex\hbox{\rlap{$\propto$}\lower 1.2ex\hbox{$\sim$}}}}
\def\deg{\ifmmode^\circ\else$^\circ$\fi}
\def\pdeg{\ifmmode $\setbox0=\hbox{$^{\circ}$}\rlap{\hskip.11\wd0 .}$^{\circ}
          \else \setbox0=\hbox{$^{\circ}$}\rlap{\hskip.11\wd0 .}$^{\circ}$\fi}
\def\arcs{\ifmmode {^{\scriptstyle\prime\prime}}
          \else $^{\scriptstyle\prime\prime}$\fi}
\def\arcm{\ifmmode {^{\scriptstyle\prime}}
          \else $^{\scriptstyle\prime}$\fi}
\newdimen\sa  \newdimen\sb
\def\parcs{\sa=.07em \sb=.03em
     \ifmmode \hbox{\rlap{.}}^{\scriptstyle\prime\kern -\sb\prime}\hbox{\kern -\sa}
     \else \rlap{.}$^{\scriptstyle\prime\kern -\sb\prime}$\kern -\sa\fi}
\def\parcm{\sa=.08em \sb=.03em
     \ifmmode \hbox{\rlap{.}\kern\sa}^{\scriptstyle\prime}\hbox{\kern-\sb}
     \else \rlap{.}\kern\sa$^{\scriptstyle\prime}$\kern-\sb\fi}
\def\ra[#1 #2 #3.#4]{#1\sup{h}#2\sup{m}#3\sup{s}\llap.#4}
\def\dec[#1 #2 #3.#4]{#1\deg#2\arcm#3\arcs\llap.#4}
\def\deco[#1 #2 #3]{#1\deg#2\arcm#3\arcs}
\def\rra[#1 #2]{#1\sup{h}#2\sup{m}}

\def\dots{\relax\ifmmode \ldots\else $\ldots$\fi}
\def\WHzsr{\ifmmode $W\,Hz\mo\,sr\mo$\else W\,Hz\mo\,sr\mo\fi}
\def\mHz{\ifmmode $\,mHz$\else \,mHz\fi}
\def\GHz{\ifmmode $\,GHz$\else \,GHz\fi}
\def\mKs{\ifmmode $\,mK\,s$^{1/2}\else \,mK\,s$^{1/2}$\fi}
\def\muKs{\ifmmode \,\mu$K\,s$^{1/2}\else \,$\mu$K\,s$^{1/2}$\fi}
\def\muKRJs{\ifmmode \,\mu$K$_{\rm RJ}$\,s$^{1/2}\else \,$\mu$K$_{\rm RJ}$\,s$^{1/2}$\fi}
\def\muKHz{\ifmmode \,\mu$K\,Hz$^{-1/2}\else \,$\mu$K\,Hz$^{-1/2}$\fi}
\def\MJysr{\ifmmode \,$MJy\,sr\mo$\else \,MJy\,sr\mo\fi}
\def\MJysrmK{\ifmmode \,$MJy\,sr\mo$\,mK$_{\rm CMB}\mo\else \,MJy\,sr\mo\,mK$_{\rm CMB}\mo$\fi}
\def\microns{\ifmmode \,\mu$m$\else \,$\mu$m\fi}

\def\muK{\ifmmode \,\mu$K$\else \,$\mu$\hbox{K}\fi}
\def\microK{\ifmmode \,\mu$K$\else \,$\mu$\hbox{K}\fi}
\def\muW{\ifmmode \,\mu$W$\else \,$\mu$\hbox{W}\fi}
\def\kms{\ifmmode $\,km\,s$^{-1}\else \,km\,s$^{-1}$\fi}
\def\kmsMpc{\ifmmode $\,\kms\,Mpc\mo$\else \,\kms\,Mpc\mo\fi}
\providecommand{\sorthelp}[1]{}

\title{B-mode polarization forecasts for \name}

\author{\small
U.~Fuskeland\inst{1}\thanks{Corresponding author: U.~Fuskeland; \url{unnif@astro.uio.no}}
\and
A.~Kaplan\inst{2}
\and
I.~K.~Wehus\inst{1}
\and
H.~K.~Eriksen\inst{1}
\and
P.~R.~Christensen\inst{3,4}
\and
S.~von Hausegger\inst{3,4,5}
\and
H.~Liu\inst{3,4,6,7}
\and
P.~M.~Lubin\inst{2}
\and
P.~R.~Meinhold\inst{2}
\and
P.~Naselsky\inst{3,4}
\and
H.~Thommesen\inst{1}
\and
A.~Zonca\inst{8}
}
\institute{\small
1. Institute of Theoretical Astrophysics, University of Oslo, Blindern, Oslo, Norway\\
2. Department of Physics, University of California, Santa Barbara, California, U.S.A.\\
3. Niels Bohr Institute, University of Copenhagen, Blegdamsvej 17, Copenhagen, Denmark\\
4. Discovery Center, University of Copenhagen, Niels Bohr Institute, Blegdamsvej 17, Copenhagen, Denmark\\
5. Department of Physics, University of Oxford, Denys Wilkinson Building, Keble Road, Oxford OX1 3RH, UK\\
6. School of Physics and Optoelectronics Engineering, Anhui
University, 111 Jiulong Road, Hefei, Anhui, China 230601. \\
7. Key Laboratory of Particle and Astrophysics, Institute of
High Energy Physics, CAS, 19B YuQuan Road, Beijing, China, 100049. \\
8. San Diego Supercomputer Center, University of California San Diego, La Jolla, U.S.A.\\
}

\begin{document}

\abstract{ We present tensor-to-scalar ratio forecasts for \name, a
  hypothetical ground-based $B$-mode experiment aiming to survey the
  cleanest regions of the Northern Galactic Hemisphere at five
  frequencies between 10 and 44~GHz. Its primary science goal would be to
  measure large-scale cosmic microwave background (CMB) polarization fluctuations at multipoles
  $\ell \lesssim 500$, and thereby constrain the primordial
  tensor-to-scalar ratio $r$. The observations for the suggested experiment would take place at the
  Summit Station ($72\degr34$N, $38\degr27$W) on Greenland, at an
  altitude of 3216 m above sea level. For this paper we simulated
  various experimental setups, and derived limits on the
  tensor-to-scalar ratio after CMB component separation using a
  Bayesian component separation implementation called Commander. When
  combining the proposed experiment with \Planck\ HFI observations for
  constraining polarized thermal dust emission, we found a projected
  limit of $r<0.02$ at 95\% confidence for the baseline
  configuration. This limit is very robust with respect to a range of
  important experimental parameters, including sky coverage, detector
  weighting, foreground priors, among others.  Overall, \name\ would have the possibility to provide deep
  CMB polarization measurements of the Northern Galactic
  Hemisphere at low frequencies.  }


\maketitle

\section{Introduction}
\label{sec:introduction}

The detection of primordial gravitational waves in the form of
large-scale polarization $B$ modes in the CMB ranks as one of the 
most important goals in current cosmology. 
To date, the two most stringent limits on the amplitude of these
waves (typically quantified by the so-called tensor-to-scalar ratio 
$r$) have been  $r<0.036$ \citep{ade:2021} and $r<0.032$ \citep{tristram:2022}, 
both at 95\% confidence.

Before BICEP2, Keck, and \Planck, the main challenge associated with
this types of measurements was simply achieving sufficient instrumental
sensitivity. The predicted amplitude of inflationary polarization
$B$ modes corresponds to less than $1\,\mu\textrm{K}$ fluctuations on
degree angular scales, and quite possibly less than $10\,$nK,
depending on the precise numerical value of $r$ \citep[see, e.g.,][and
  references therein]{lyth:1999}. For this reason, the latest
generation of $B$-mode instruments typically deploy hundreds or
thousands of detectors, simply in order to achieve sufficient raw
instrumental sensitivity \citep[e.g.,][]{austermann:2012,bicep2:2014,inoue:2016,thornton:2016}.

However, after BICEP2, Keck, and \Planck, this situation has changed
dramatically \citep{bicep2planck:2015}. Because of so-called
foreground contamination in the form of polarized synchrotron and
thermal dust emission from the Milky Way \citep{planckX:2016}, it has
now become abundantly clear that raw sensitivity alone no longer
suffices in order to improve existing constraints by a significant
factor. The focus has rather turned to removing the foreground
contaminants to a precision better than the noise level. This topic is
called CMB component separation, and it is one of the central research
topics in CMB physics today.

In order to reliably distinguish between cosmological and
astrophysical sky signals, two main issues are of particular
importance when designing a new experiment, namely field selection and
frequency coverage. First, as illustrated by the \Planck\
polarization maps \citep{planckX:2016}, the amplitude of polarized
Galactic foregrounds varies by orders of magnitudes over the sky. In
order to make the overall analysis problem as simple as possible, it
is wise to observe the cleanest possible regions on the sky
\citep{planckXXX:2016}. Of course, this strategy has been adopted by
most experiments of this type deployed to date; however, before \Planck, only
very crude estimates of the amplitude of polarized thermal dust had
been available. After \Planck, we have a much more complete picture of
the situation. And in this new situation, an interesting question has
emerged: while most previous experiments have targeted the cleanest
region of the Southern Galactic Hemisphere, there are possible hints
in the latest \Planck\ measurements that selected regions of the
Northern Galactic Hemisphere could exhibit even lower overall
foreground contamination \citep{planckX:2016,planckXXX:2016}. If this
is correct, then it is clearly of great interest to deploy
high-sensitivity experiments in the Northern Hemisphere in the near
future, to supplement the data from, for instance, QUIJOTE \citep{quijote:2010}, C-BASS \citep{jones:2018}, S-PASS \citep{caretti:2019}, or even dedicated SKA surveys \citep{Basu:2019xrm}.

\begin{table*}[t]                                                    
\begingroup                                                          
\newdimen\tblskip \tblskip=2pt
\caption{Summary of instrument properties (top section), model
  parameters (middle section), and experimental setups (bottom section). 
  Temperatures are given in thermodynamic units, $\mu\mathrm{K}_\mathrm{CMB}$. \label{tab:instrument} }
\nointerlineskip               
\vskip -3mm
\footnotesize                                                       
\setbox\tablebox=\vbox{                 
\newdimen\digitwidth                
\setbox0=\hbox{\rm 0}
\digitwidth=\wd0
\catcode`*=\active
\def*{\kern\digitwidth}
\newdimen\signwidth
\setbox0=\hbox{+}
\signwidth=\wd0
\catcode`!=\active
\def!{\kern\signwidth}
\newdimen\decimalwidth
\setbox0=\hbox{.}
\decimalwidth=\wd0
\catcode`@=\active
\def@{\kern\signwidth}
\halign{ \hbox to 3.5in{#\leaderfil}\tabskip=1.0em&
    \hfil#\hfil\tabskip=1em&
    \hfil#\hfil\tabskip=1em&
    \hfil#\hfil\tabskip=1em&
    \hfil#\hfil\tabskip=1em&
    \hfil#\hfil\tabskip=1em&
    \hfil#\hfil\tabskip=1em&
    #\hfil\tabskip=0pt\cr
\noalign{\doubleline}
Frequency (GHz) & 10 & 15 & 20 & 30 & 44 & 90 & 143 \cr
\noalign{\vskip 3pt\hrule\vskip 8pt}
\omit {\bf Detector specification}\hfil \cr
\noalign{\vskip 2pt}
\hglue 1em\ Detector type&HEMT&HEMT&HEMT&HEMT&Bolo&Bolo&Bolo\cr
\hglue 1em\ RMS per horn ($\mu\mathrm{K}\sqrt{\mathrm{s}}$)&316&316&433&361&200&300&300\cr
\hglue 1em\ Horns per telescope&7&13&19&25&40&160&320\cr
\hglue 1em\ NET per telescope ($\mu\mathrm{K}\sqrt{\mathrm{s}}$)&120&88&102&72&16&24&17\cr
\hglue 1em\ Mean ($Q$,$U$) noise per $1^{\circ}$ pixel$^a$ ($\mu\mathrm{K}$)&5.1&2.1&2.0&1.3&0.3&0.4&0.3\cr
\hglue 1em\ FWHM beam size (arcmin)&80.0&53.3&40.0&26.7&18.18&8.89&5.59\cr
\noalign{\vskip 3pt\hrule\vskip 8pt}
\multispan{8}{\bf Model summary} \hfil\cr
\noalign{\vskip 2pt}
\hglue 1em\ Mean ($P/\sqrt{2}$) total signal$^a$ ($\mu\mathrm{K}$)&279&80&33&10&3.7&2.5&5.2\cr
\hglue 1em\ Mean ($P/\sqrt{2}$) Thermal dust$^a$ ($\mu\mathrm{K}$)&0.07&0.12&0.18&0.34&0.61&2.0&5.0\cr
\hglue 1em\ Mean ($P/\sqrt{2}$) Synchrotron$^a$ ($\mu\mathrm{K}$)&279&80&33&9.4&3.0&0.38&0.12\cr
\hglue 1em\ RMS (Q) CMB$^a$ ($\mu\mathrm{K}$)&0.43&0.43&0.43&0.43&0.43&0.43&0.43\cr
\hglue 1em\ Thermal dust scaling factor&0.007&0.013&0.020&0.038&0.068&0.22&0.55\cr
\hglue 1em\ Synchrotron scaling factor&13.3&3.8&1.56&0.45&0.14&0.02&0.006\cr
\hglue 1em\ Spinning dust scaling factor&0.52&0.44&0.30&0.10&0.026&$2\cdot10^{-4}$&$1\cdot10^{-9}$\cr
\noalign{\vskip 3pt\hrule\vskip 8pt}
\multispan{8}{\bf Integration time per channel for various experimental setups
  in telescope-years} \hfil\cr
\noalign{\vskip 2pt}
\hglue 1em\ \emph{SN10} --- S/N weights, $10^{\circ}$ radius disk&0.2&0.3&1.5&6&8&0&0\cr
\hglue 1em\ \emph{SN20} --- S/N weights, $20^{\circ}$ radius disk&0.2&0.3&1.5&6&8&0&0\cr
\hglue 1em\ \emph{SN30} --- S/N weights, $30^{\circ}$ radius disk&0.2&0.3&1.5&6&8&0&0\cr
\hglue 1em\ \emph{SN65} --- S/N weights, $65^{\circ}$ radius disk&0.2&0.3&1.5&6&8&0&0\cr
\hglue 1em\ \emph{U10} --- uniform weights, $10^{\circ}$ radius disk&2&2&2&2&8&0&0\cr
\hglue 1em\ \emph{SN10H} --- S/N weights with 90+143 bolo, $10^{\circ}$ radius disk&0.1&0.2&0.7&3&4&4.5&0.5\cr
\noalign{\vskip 2pt\hrule\vskip 2pt}
}}
\endPlancktablewide
\tablenote {{\rm a}} Evaluated for the baseline scanning strategy with
$10^{\circ}$ opening angle.\par
\endgroup
\end{table*}

The second issue concerns frequency coverage. Traditionally, most
high-sensitivity experiments have employed bolometers above 90 GHz,
primarily because of their excellent white noise performance and small
focal plane area footprint; it is simply easier to achieve a high
sensitivity at these frequency channels. 
For a long time, the astrophysics complexity of the sky as observed at the
corresponding frequencies was of a secondary concern.
However, the latest \Planck\ measurements have shown that the
amplitude of polarized thermal dust, which dominates on frequencies
above 70~GHz, is higher than what many expected only a few years
prior \citep{bicep2:2014,bicep2planck:2015,planckX:2016}.

In this new ``high-foreground'' situation, many important questions
are currently being addressed quantitatively within the community. One example addresses whether thermal dust emission exhibits a simpler or more complex morphology than synchrotron emission in the presence of small-scale turbulent magnetic fields \citep[e.g.,][]{cho:2002,hausegger:2019,kim:2019}.
Another question is whether the steep spectral index of synchrotron emission ($T_{\nu} \propto \nu^{-3}$) is helpful or detrimental for component separation purposes, compared to the flatter index of thermal dust emission ($T_{\nu} \propto \nu^{1.5}$); a steeper index makes synchrotron emission inherently less degenerate with the CMB than thermal dust emission, since the CMB spectrum itself is flat. On the other hand, the steep index also implies larger absolute amplitudes at low frequencies. On the low frequency side, there have also been efforts to investigate whether the synchrotron emission follows a power law throughout the observed frequency range, or whether it should contain a curvature \citep[e.g.,][]{kogut:2012,delahoz:2023}. 
A third question is whether synchrotron or thermal dust emission correlate more strongly with the CMB signal for a given experimental setup; this can for instance happen if the frequency range is insufficient to properly fit a given component. Lastly, possible frequency decorrelation may be induced by the 3D spatial variation of the foreground properties within the Milky Way, which, if not properly taken into account, can leak into the CMB \citep[e.g.,][]{tassis:2015,chluba:2017,vacher:2022,vacher:2023,azzoni:2021,azzoni:2023,mangilli:2021}.

The precise answers to these questions depend sensitively on what the
true sky looks like, both at low and high frequencies. Unfortunately,
while \Planck\ has provided an excellent view of frequencies higher than
100\,GHz, and both QUIJOTE and C-BASS provide strong constraints
between 5 and 19\,GHz, an even higher sensitivity is still warranted.  In order to make
robust progress on all of the above questions, it is therefore
essential to deploy new high-sensitivity low-frequency experiments
targeting large regions of the cleanest parts of the sky.

In the following, we describe a potential experiment aiming to do
this, called \name. If funded, this experiment will deploy five
receivers between 10 and 44\,GHz on multiple low-cost telescopes, with
a map-level sensitivity that is about four times higher than the published QUIJOTE multifrequency instrument (MFI) maps in the range $10-20$~GHz \citep{quijote:2023}. 
We note however that QUIJOTE will also have a new version of the low frequency instrument called MFI2 with a plan to improve the sensitivity by a factor of 2 or 3.
\name\ will be based at the Summit Station at Greenland, at an altitude of
3216 m above sea level. This site offers several unique and
attractive features, as described in greater detail in the next
section, but for the purposes of the current discussion the most
important is simply that it provides a direct and continuous view of
the cleanest regions of the Northern Galactic Hemisphere.

A similar experiment which also targets the Northern Galactic Hemisphere has been examined in \cite{araujo:2014}, where the instrument is a compact CMB polarimeter based on lumped-element kinetic inductance detectors (LEKIDs). Observations would in this case be made from Isi Station in Greenland, and at higher frequencies than \name\, namely 150, 210 and 267\,GHz.

The goal of the current paper is to provide preliminary forecasts for
\name\ in terms of effective limits on the tensor-to-scalar
ratio. These forecasts are derived from ideal and simplified
simulations that account for foregrounds and white noise only, but not
instrumental systematics.
The rest of the paper is organized as follows: In
Sect. \ref{sec:experiment}, we provide a more detailed overview of the
technical specifications of the proposed experiment and the site, and discuss
various aspects related to possible scanning strategies. In Sect.
\ref{sec:method}, we describe the methodology used to produce the
forecasts, both in terms of simulations and posterior mapping
framework. In Sect. \ref{sec:results} we summarize our main results,
before concluding in Sect. \ref{sec:conclusions}.

\section{\name\ overview}
\label{sec:experiment}

The \name\ concept relies on combining existing and proven detector
technology with a low-cost telescope design \citep{childers:2005}. 
The top section of Table~\ref{tab:instrument} provides
an overview of the default detector combinations that are currently
being considered. For frequencies between 10 and 30~GHz, coherent detectors such as those using High Electron Mobility Transistor amplifiers (HEMTs) may be preferred as they provide a path to mitigation of the rapidly increasing earth and orbitally based Radio Frequency Interference sources \citep{barron:2022, hoyland:2022}
while at higher frequencies, low-noise bolometers (e.g.,
transition edge sensor (TES) bolometers) are preferred. The quoted
HEMT noise equivalent temperature (NET) levels represent real values measured on the sky with
in-hand detectors, while the bolometer values represent typical values 
obtained by similar experiments \citep{kermish:2012,george:2012,thornton:2016,ade:2015}. 

\begin{figure}[t]
\begin{center}
  \mbox{\includegraphics[width=\linewidth,clip=]{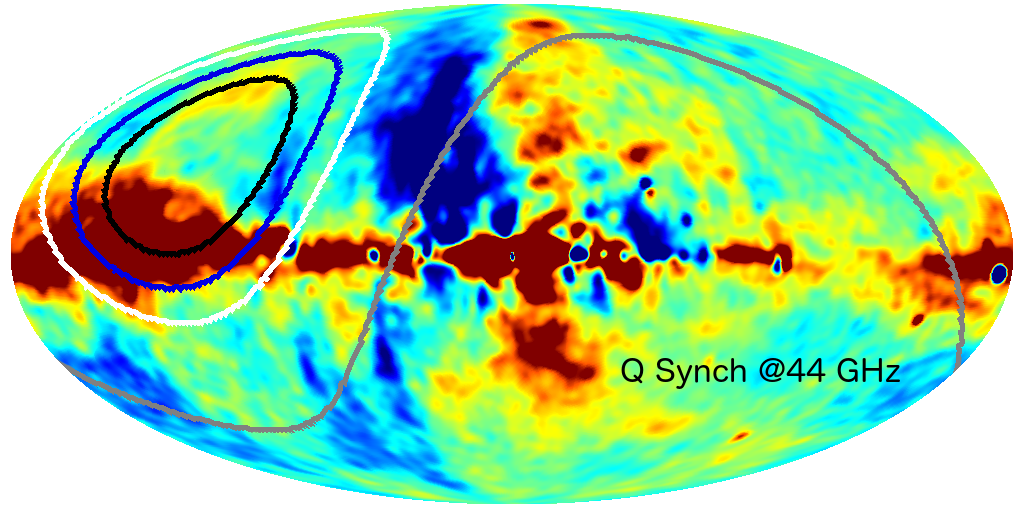}}\\
  \mbox{\includegraphics[width=\linewidth,clip=]{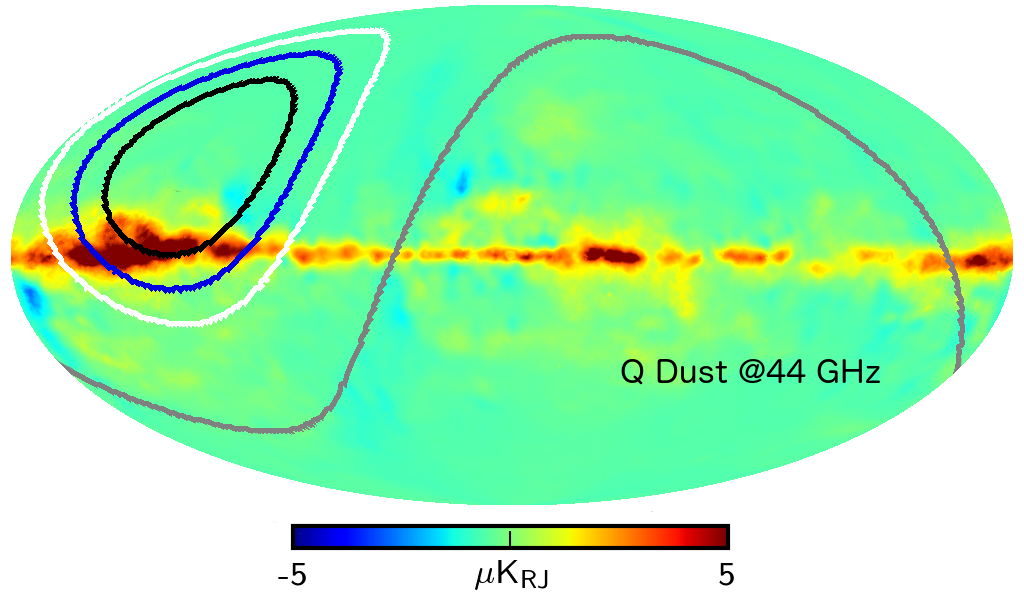}}
\end{center}
\caption{Location of the \name\ fields superimposed on the
  \Planck\ Stokes $Q$ synchrotron emission map (top panel) and
  the \Planck\ Stokes $Q$ thermal dust emission map (bottom 
  panel), both evaluated at 44~GHz. The lines show the location of
  the $10^\circ$ field (black), the $20^\circ$ field (blue), the
  $30^\circ$ field (white) and the $65^\circ$ field
  (gray). 
}
\label{fig:greenlandpatch}
\end{figure}

The \name\ optical system would be based on recently developed carbon
fiber mirror technology developed at the University of California
Santa Barbara (UCSB), allowing for both cheap and lightweight production of
multiple telescopes \citep{childers:2005}. The numbers
of horns per telescope listed in Table~\ref{tab:instrument} correspond
to an optical system with a 2.2 m primary and 0.9 m secondary mirror,
allowing for 7 horns at 10~GHz, 40 horns at 44~GHz, and 320 horns at
143~GHz. The effective angular resolution is $18.2'$ full width half 
maximum (FWHM) at 44~GHz.

As already noted, one of the most unique features of \name\ is its
deployment location at the Summit Station ($72\degr34$N, $38\degr27$W)
on Greenland \citep{andersen:2006}. At a latitude of $72\degr$N, this
site provides unique access to the Northern Galactic Hemisphere,
which hitherto has been largely unexplored by suborbital
high-sensitivity CMB experiments. Summit Station is also an observing 
site with potentially optimal seeing conditions in the winter, particularly with respect to 
precipitable water vapor \citep{suen:2014}, though it was found that summer conditions present significant weather challenges.

The basic scanning strategy envisioned for \name\ is based on full
revolution constant elevation scans. The boresight of the telescope
swipes out a circle of constant elevation in horizon coordinates at
any given time, while the sky rotation slowly moves this circle across
the sky. The most important free parameter in this scanning strategy
is the co-elevation (or opening) angle $\alpha$, which is the angle
between the zenith and the boresight; the larger this angle is, the
bigger sky fraction will be observed. In this paper, we consider
four specific choices of this angle, namely $\alpha=10^{\circ}$,
$20^{\circ}$, $30^{\circ}$ and $65^{\circ}$, but we note that the real
experiment will most likely consider a wide range of such angles, for
instance observing one day at each elevation, in order to uniformize
the overall scan pattern. Indeed, the $\alpha=65^{\circ}$ presented
below represents precisely such a union of multiple co-elevation
angles, and the actual value of $65^{\circ}$ simply quantifies the
largest angle in the observation set. 
Figure~\ref{fig:greenlandpatch} shows outlines of the sky coverages for
each of the above co-elevation angles, overplotted on the \Planck\
synchrotron (top) and thermal dust (bottom) Stokes $Q$ polarization
amplitude maps, evaluated at 44 GHz. A latitude of
$72^{\circ}$N allows for continuous observations of a single field on
the sky, although due to the constant elevation scan there is a hole in the field at the celestial pole, which we for visualization purposes have dropped in this figure.

Another powerful advantage of observations from a latitude of
$72^{\circ}$N is the unique cross-linking properties, that is, multiple crossings over the same pixel from different orientations on the sky during a long integration scan. This is used to remove offsets estimated by minimizing the scatter among different measurements of each pixel and is extremely powerful for suppressing a wide range of instrumental systematic effects that are important for polarization measurements.
The basic scanning strategy is identical to what was proposed in 
\cite{araujo:2014}, and Fig.~8 (left panel) in that paper
shows the scan path for 10
days of observations, selecting a different co-elevation angle each day. 

In the summer of 2018 UCSB deployed a 10\,GHz receiver with a 2.2\,m primary mirror to Summit Station using a foldable design which allowed for storage in a standard ISO shipping container. This successfully demonstrated a method of shipping and quickly deploying a telescope in an extremely remote location, as well as the ability to quickly stow and redeploy the instrument in the face of changing weather conditions. The forecasting results show the potential for a larger experiment of this type with multiple telescopes and observing frequencies, which is made more feasible by the ability to ship and rapidly deploy fully functional telescopes to the desired observation site. 

The primary goals of the mission were to assess the viability of the site for cosmological observations, as well as to conduct preliminary measurements and detect the Stokes $Q$ and $U$ components of the microwave foregrounds. 
However, the climate conditions during the summer deployment turned out to be more challenging than anticipated, with extended periods of overcast sky, freezing fog, wind, and blowing snow. Winter conditions at Summit Station may well be substantially better for observations, but this would need to be demonstrated on the ground before a full deployment. Notably, the observing strategy and analysis presented here do not require use of Summit Station, and there are other high quality Northern Hemisphere sites available \citep{marvil:2006,suen:2014}.

\section{Forecast methodology}
\label{sec:method}

The goal of the current paper is to quantify the performance of an
experiment such as \name\ in terms of overall $B$-mode polarization
sensitivity after component separation under ideal conditions. For
this task, we combine simple simulations of the astrophysical sky with
a well-established Bayesian component separation framework called
Commander \citep{eriksen:2004,eriksen:2008}, and compare a range of
different experimental setups in terms of their effective constraints on
the tensor-to-scalar ratio.

\subsection{Simulations}
\label{sec:simulations}

For our simulations, we adopt a simple model consisting of four
components, namely CMB, synchrotron, and thermal dust emission, as well
as instrumental noise. We model synchrotron emission with a simple
power-law spectral energy density (SED) and thermal dust emission by a
one-component modified blackbody SED \citep{planckX:2016}. We adopt
thermodynamic temperature units throughout, and this simple model then
takes the following form,
\begin{align}
  d_{\nu}(p) =& s(p) + \\
  &a_{\mathrm{s}}(p)
  \left(\frac{\nu}{\nu_\mathrm{s}}\right)^{\beta_{\mathrm{s}}(p)}\gamma(\nu) + \\
  &a_{\mathrm{d}}(p)
  \left(\frac{\nu}{\nu_\mathrm{d}}\right)^{\beta_{\mathrm{d}}(p)+1}
  \frac{e^{\frac{h\nu_{\mathrm{d}}}{kT_{\mathrm{d}}}}-1}{e^{\frac{h\nu}{kT_{\mathrm{d}}}}-1}\gamma(\nu)
  +\\
&n_{\nu}(p),
\label{eq:model0}
\end{align}
where $d_{\nu}(p)$ is the total simulated sky map at frequency $\nu$
and pixel $p$; $s$ is the CMB signal; $a_{\mathrm{s}}$ and
$\beta_{\mathrm{s}}$ are the synchrotron amplitude (at a reference
frequency of $\nu_\mathrm{s}=23\,\textrm{GHz}$) and spectral index;
$a_{\mathrm{d}}$, $\beta_{\mathrm{d}}$ and $T_{\mathrm{d}}$ are the
thermal dust amplitude (at a reference frequency of
$\nu_\mathrm{d}=353\,\textrm{GHz}$), spectral index and temperature,
respectively; $\gamma(\nu)$ is the conversion factor between
Rayleigh-Jeans and thermodynamic temperatures for differential
observations\footnote{$\gamma(\nu)=\frac{x^2 e^{x}}{(e^{x}-1)^2}$
  where $x=h\nu/k_\textrm{B}T_{\mathrm{CMB}}$; $h$ is Planck's
  constant, and $k_{\mathrm{B}}$ is Boltzmann's constant}; and $n$ is
instrumental noise.

Simulated CMB sky maps are generated using the
HEALPix\footnote{http://healpix.jpl.nasa.gov} \citep{gorski:2005}
facility called \texttt{synfast}, combined with a theoretical
$\Lambda$CDM angular power spectrum computed with CAMB
\citep{lewis:2000} for various values of the tensor-to-scalar ratio
and an optical depth of reionization of $\tau=0.06$. All other
cosmological parameters are fixed at best-fit \Planck\ values
\citep{planckXIII:2016}. 

For the synchrotron component, we adopt the Wilkinson Microwave Anisotropy Probe (\WMAP) 
K-band (23~GHz) Stokes $Q$ and $U$ maps
as a spatial template \citep{bennett:2013}, and a spatially constant
spectral index of $\beta_{\mathrm{s}}=-3.1$ across our
fields. Likewise, for thermal dust emission we adopt the \Planck\
353~GHz Stokes $Q$ and $U$ maps as a spatial template \citep{planckI:2016}, and
spatially constant values of $\beta_{\mathrm{d}}=1.5$ and
$T_{\mathrm{d}}=21\,\mathrm{K}$ \citep{planckX:2016}. We note that while
we do adopt spatially constant spectral indices in the simulation, no
such assumptions are enforced or assumed in the analysis, but both
$\beta_{\mathrm{s}}$ and $\beta_{\mathrm{d}}$ are fitted independently
pixel-by-pixel. The thermal dust temperature is kept fixed on the input 
value in the fit, since this parameter is not constrained by the frequencies in
question for \name.

As a test of potential modeling errors, we also consider one case that
includes polarized spinning dust in the input simulation, but not in
the fitted model. The SED for the spinning dust component is taken to
match the corresponding intensity SED for spinning dust as reported in
Fig.~51 of \cite{planckX:2016}, while the \Planck\ 353 GHz thermal dust
polarized amplitude map is adopted as a spatial template. Finally, we
adopt a polarization fraction of 1\%, consistent with current upper
limits \citep[e.g.,][]{quijote:2015}.

All input sky maps are downgraded to $N_\mathrm{side}=64$ and 
smoothed to a common resolution of $1^{\circ}$
FWHM before co-addition. On top of this, we add Gaussian white noise
modulated by the scanning strategy, with a root mean square (RMS) per pixel given by
$\sigma_0/\sqrt{N_{\mathrm{obs}}(p)}$, while properly accounting for
smoothing; $N_{\mathrm{obs}}(p)$ is the number of observations in
pixel $p$, computed by simulating the scan path of the instrument for
one year of observations. The noise is assumed independent in Stokes
$Q$ and $U$.

For each considered experimental setup, we analyze 20 independent
simulations with different CMB and noise realizations. All quoted
results, unless specified otherwise, correspond to averages over these
20 realizations.

\subsection{Posterior mapping}
\label{sec:commander}

We extract an estimate of the tensor-to-scalar posterior distribution,
$P(r|d)$, using a Bayesian analysis framework called Commander
\citep{eriksen:2004,eriksen:2008}. Throughout this paper, we use Commander1 \citep{eriksen:2008}, 
which is very well suited when it comes to exploring a diversity of cases like in this paper. 
This machinery has already been
used extensively for similar types of analyses in the literature, both
for real \citep[e.g.,][]{planckX:2016} and simulated data
\citep[e.g.,][]{armitage-caplan:2012,remazeilles:2016,fuskeland:2023}. 
The latest Commander implementation, Commander3 \citep{galloway:2023}, is able to perform end-to-end analysis all the way from time ordered data to cosmological parameters, as demonstrated by the BeyondPlanck \citep{bp:2023} and Cosmoglobe \citep{watts:2023} projects. However, this implementation also has a much higher computational cost than Commander1, and is not yet suitable for fast exploration of many experimental configurations as considered in this paper.
In the following, we provide a brief summary of the main ideas, and refer
the interested reader to the dedicated papers for full details.

In short, Commander implements a specific Markov Chain Monte Carlo
sampling algorithm for parametric component separation. In this
approach, the first step is to map out the full joint posterior
including all sky components, $P(s, a_{\mathrm{s}},
\beta_{\mathrm{s}}, a_{\mathrm{d}}, \beta_{\mathrm{d}}|d)$, and then
constrain the tensor-to-scalar ratio posterior from the corresponding
marginal CMB sky map posterior. The first step in this process is
performed by a special technique called Gibbs sampling
\citep[e.g.,][]{gelman:2013}, in which samples from the full posterior
distribution are drawn by iteratively sampling from all relevant
conditional distributions. For the purposes of the current paper, this
can be summarized in the following three-step Gibbs sampler,
\begin{align}
(s,a_{\mathrm{s}},a_{\mathrm{d}})^{i+1}   &\leftarrow
P(s, a_\mathrm{s}, a_\mathrm{d} |d, \beta_{\mathrm{s}}^{i},
\beta_{\mathrm{d}}^{i})  \\
\beta_{\mathrm{s}}^{i+1}   &\leftarrow
P(\beta_{\mathrm{s}}|d, s^{i+1}, a_{\mathrm{s}}^{i+1},
a_{\mathrm{d}}^{i+1}, \beta_{\mathrm{d}}^{i})  \\
\beta_{\mathrm{d}}^{i+1}   &\leftarrow
P(\beta_{\mathrm{d}}|d, s^{i+1}, a_{\mathrm{s}}^{i+1}, \beta_{\mathrm{s}}^{i+1},
a_{\mathrm{d}}^{i+1}),
\end{align}
where $\leftarrow$ means sampling from the expression on the right
hand side.  For a full account of the sampling steps, see
\cite{eriksen:2008}. In short, we first sample all linear degrees of
freedom (i.e., component amplitudes) jointly from a multivariate
Gaussian, while all nonlinear parameters (i.e., spectral indices) are
sampled iteratively by an explicit inversion sampler.

We run this iterative sampler until convergence, which typically
requires $O\mathcal(10^5)$ samples for robust convergence. From the
resulting CMB sky map samples we then compute the posterior mean CMB
map, $\hat{\mathbf{s}}$, and effective noise covariance matrix,
$\mathbf{N}$, with elements
\begin{align}
  \hat{s}(p) &= \left<s^i(p)\right>\\
  N_{pp'} &= \left< (s^i(p)-\hat{s}(p)) (s^i(p')-\hat{s}(p'))\right>,
\end{align}
where brackets indicate average over samples. Finally, these sample
averaged objects are fed into a standard Gaussian likelihood of the
form
\begin{equation}
\mathcal{L}(r) \propto \frac{e^{-\frac{1}{2} \hat{\mathbf{s}}^t
    \left(\mathbf{S}(r) +
    \mathbf{N}\right)^{-1}\hat{\mathbf{s}}}}{\sqrt{|\mathbf{S}(r) +
    \mathbf{N}|}},
\label{eq:lnL}
\end{equation}
where $\mathbf{S}(r)$ is the CMB signal covariance matrix evaluated
for the appropriate value of $r$ and the best-fit $\Lambda$CDM
parameters described above. We adopt a uniform prior for all positive
values of $r$, and the posterior is therefore numerically equal to the
likelihood in Eq.~\eqref{eq:lnL}. To map out this posterior, we simply
map out Eq.~\eqref{eq:lnL} over a one-dimensional grid in $r$. 

\begin{figure*}[p]
\vspace{-0.05in}
\centering
\subfigure{\includegraphics[width=0.208\linewidth,clip=true,trim=0.3in 0in 0.05in 0.1in]{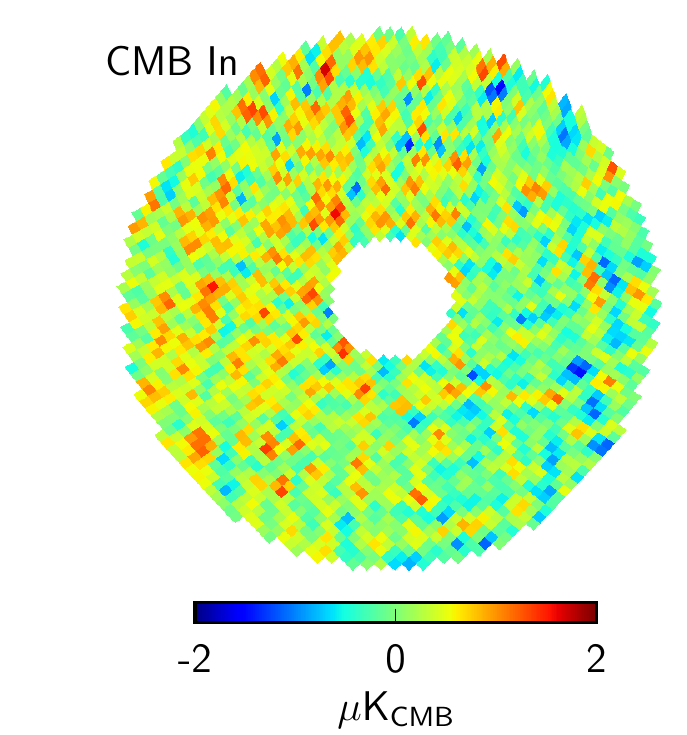}}
\subfigure{\includegraphics[width=0.208\linewidth,clip=true,trim=0.3in 0in 0.05in 0.1in]{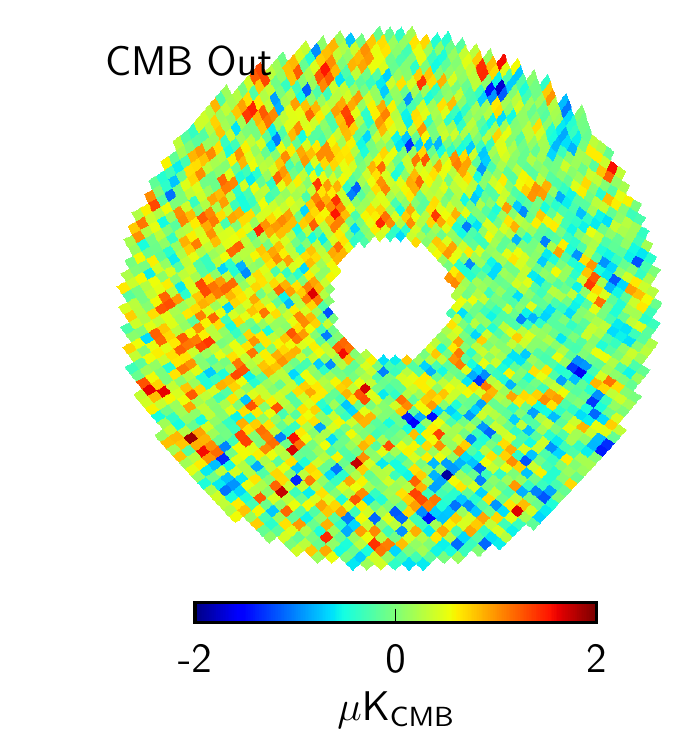}}
\subfigure{\includegraphics[width=0.208\linewidth,clip=true,trim=0.3in 0in 0.05in 0.1in]{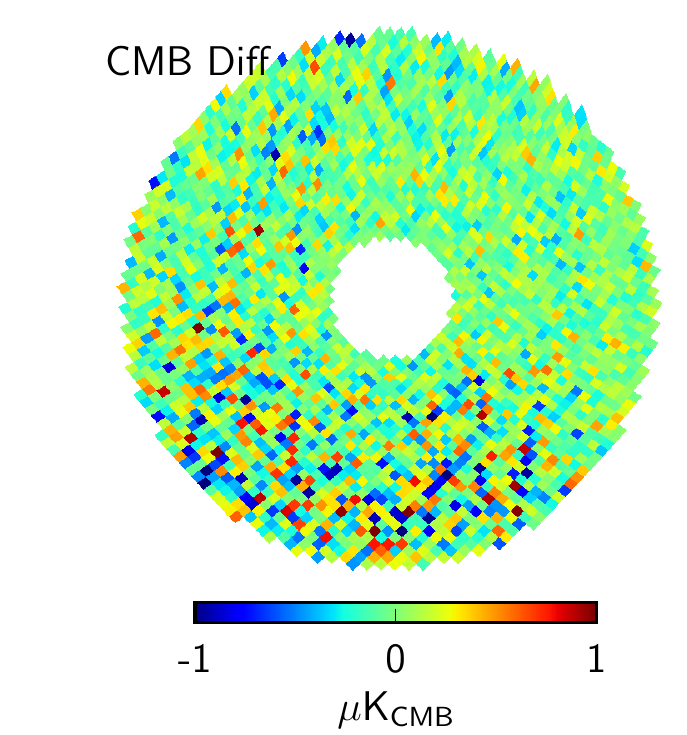}}
\subfigure{\includegraphics[width=0.208\linewidth,clip=true,trim=0.3in 0in 0.05in 0.1in]{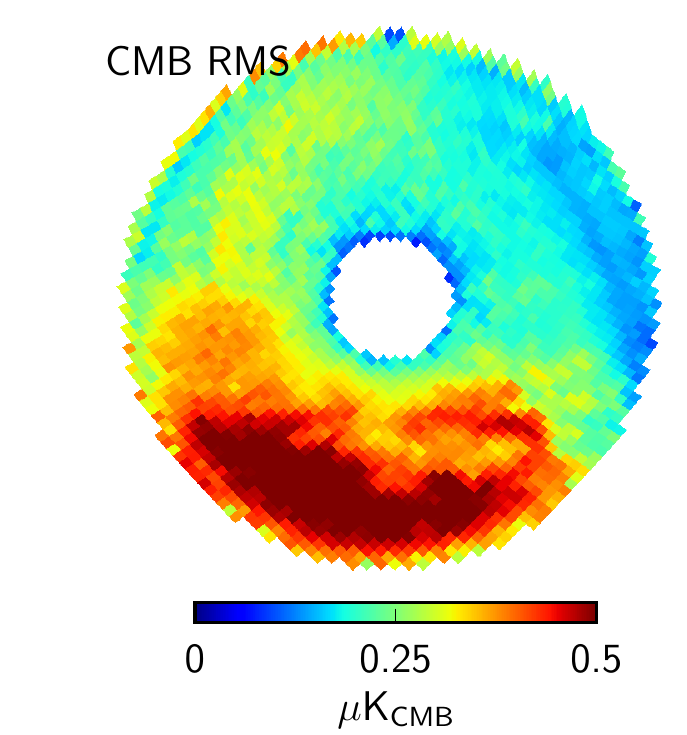}}

\subfigure{\includegraphics[width=0.208\linewidth,clip=true,trim=0.3in 0in 0.05in 0.1in]{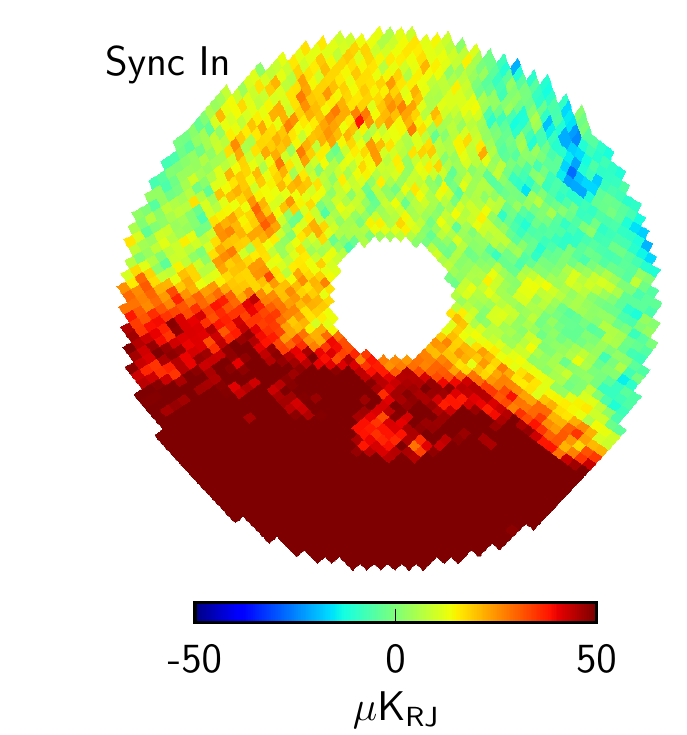}}
\subfigure{\includegraphics[width=0.208\linewidth,clip=true,trim=0.3in 0in 0.05in 0.1in]{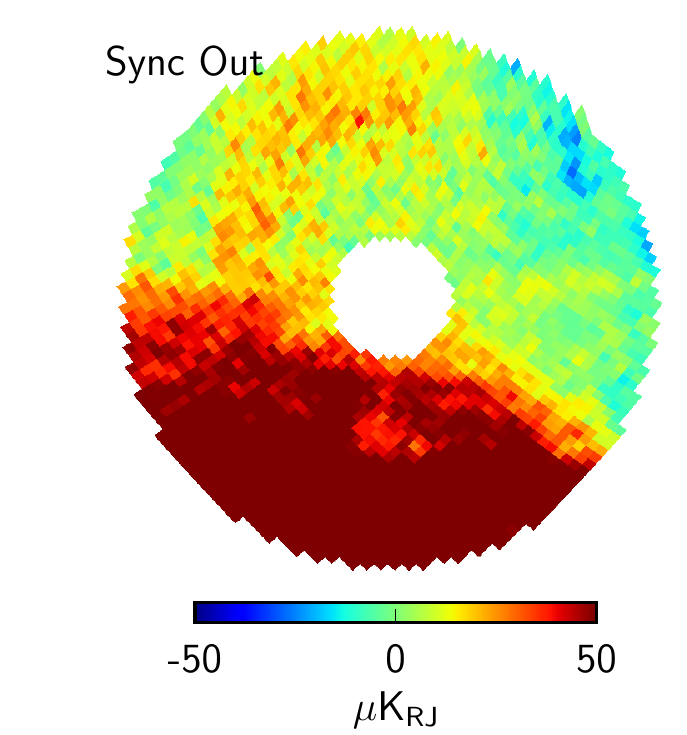}}
\subfigure{\includegraphics[width=0.208\linewidth,clip=true,trim=0.3in 0in 0.05in 0.1in]{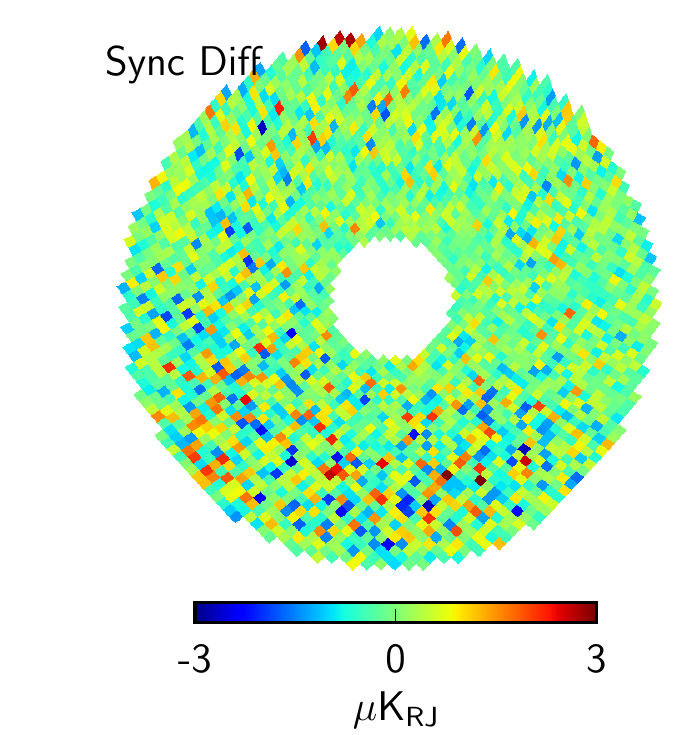}}
\subfigure{\includegraphics[width=0.208\linewidth,clip=true,trim=0.3in 0in 0.05in 0.1in]{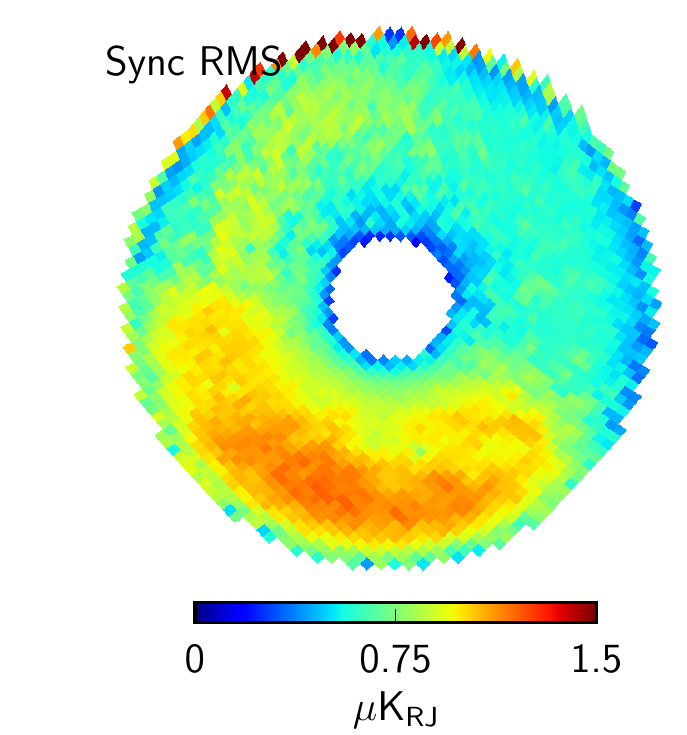}}

\subfigure{\includegraphics[width=0.208\linewidth,clip=true,trim=0.3in 0in 0.05in 0.1in]{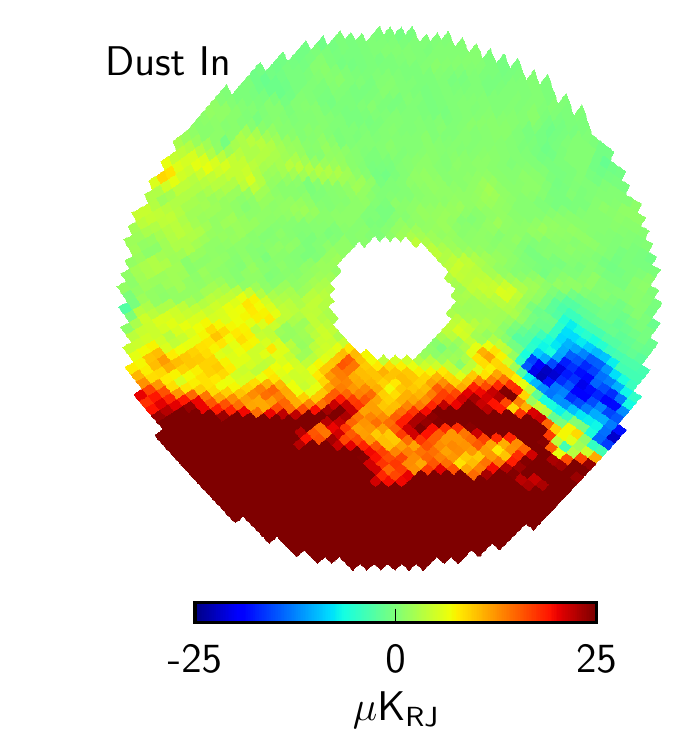}}
\subfigure{\includegraphics[width=0.208\linewidth,clip=true,trim=0.3in 0in 0.05in 0.1in]{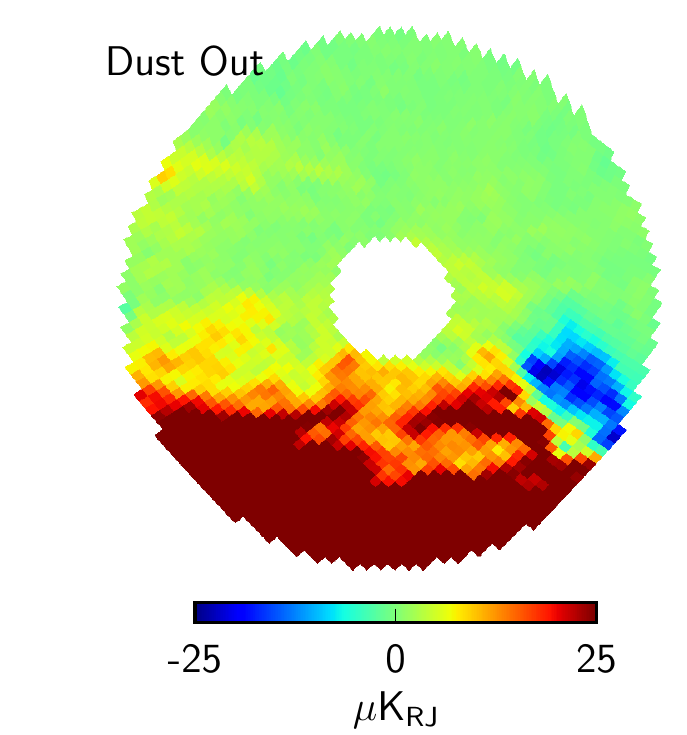}}
\subfigure{\includegraphics[width=0.208\linewidth,clip=true,trim=0.3in 0in 0.05in 0.1in]{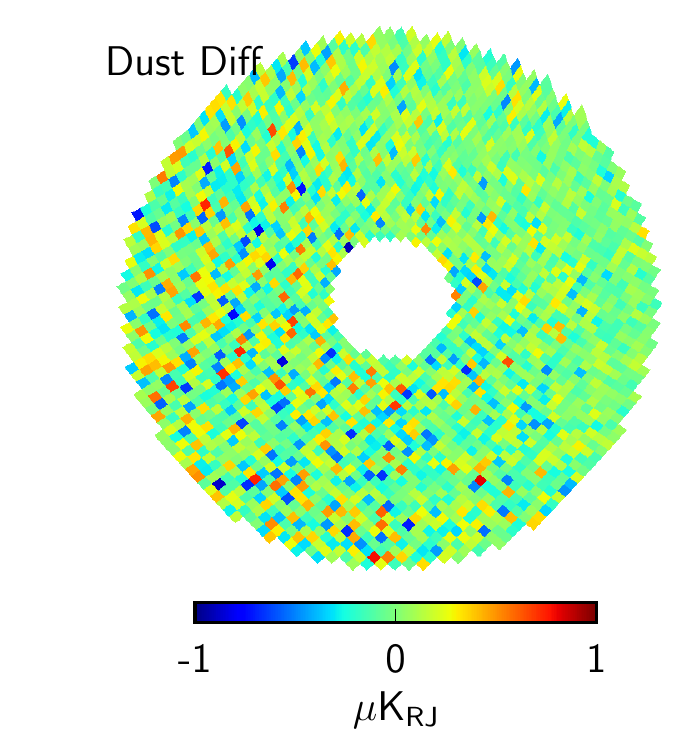}}
\subfigure{\includegraphics[width=0.208\linewidth,clip=true,trim=0.3in 0in 0.05in 0.1in]{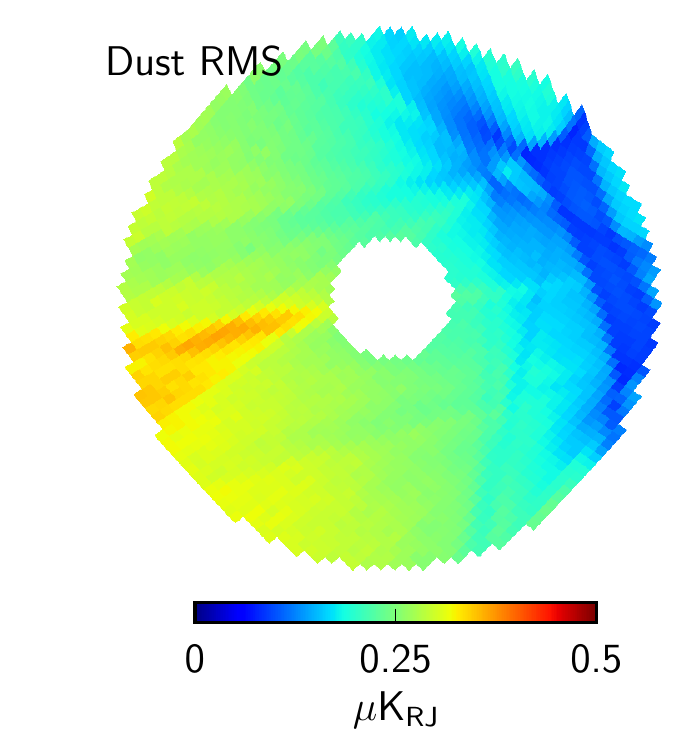}}

\subfigure{\includegraphics[width=0.208\linewidth,clip=true,trim=0.3in 0in 0.05in 0.1in]{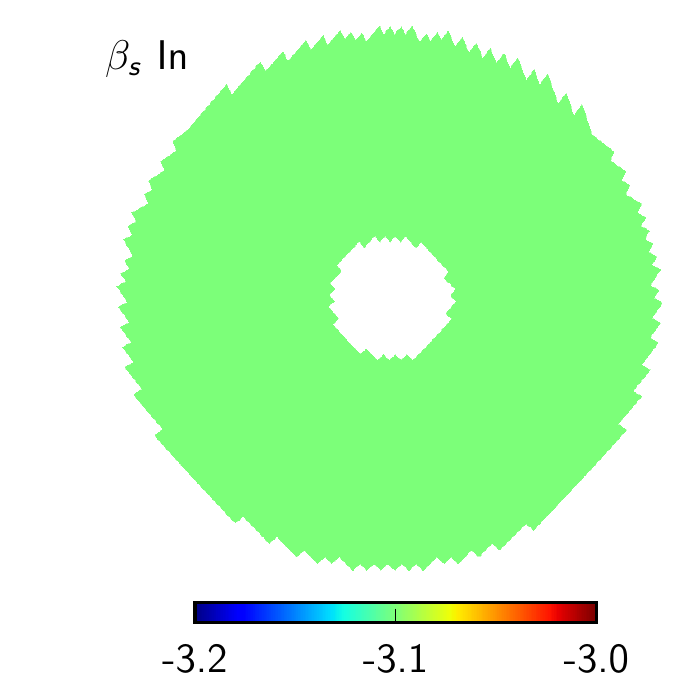}}
\subfigure{\includegraphics[width=0.208\linewidth,clip=true,trim=0.3in 0in 0.05in 0.1in]{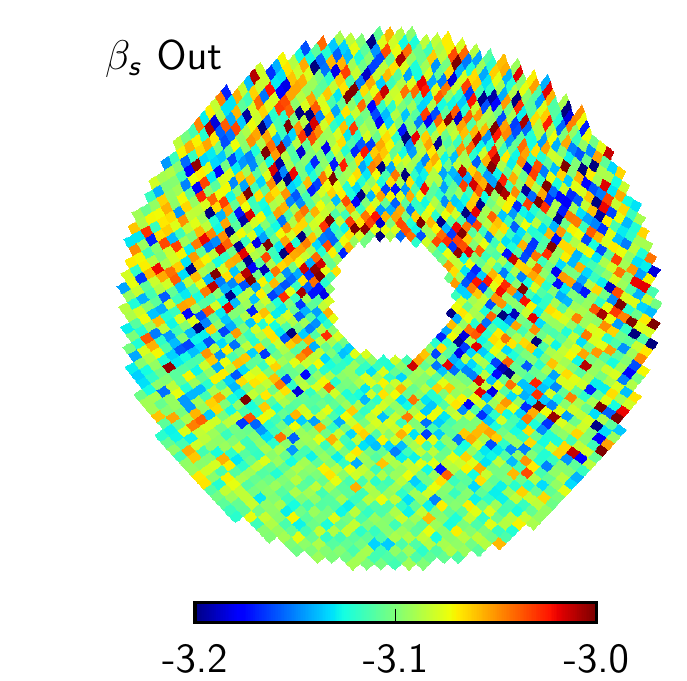}}
\subfigure{\includegraphics[width=0.208\linewidth,clip=true,trim=0.3in 0in 0.05in 0.1in]{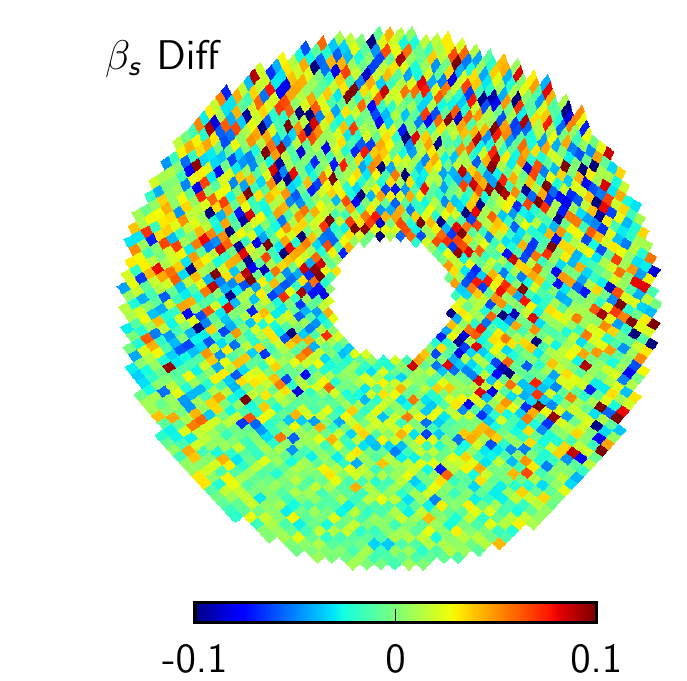}}
\subfigure{\includegraphics[width=0.208\linewidth,clip=true,trim=0.3in 0in 0.05in 0.1in]{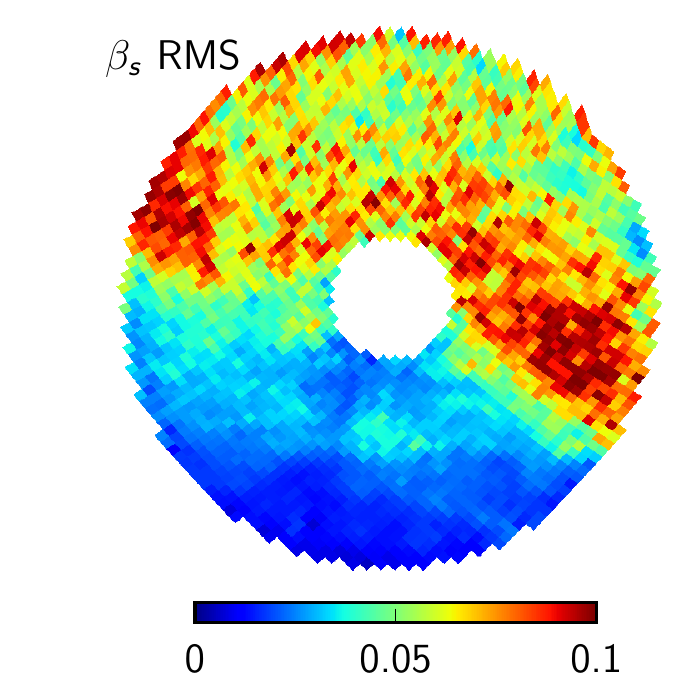}}

\subfigure{\includegraphics[width=0.208\linewidth,clip=true,trim=0.3in 0in 0.05in 0.1in]{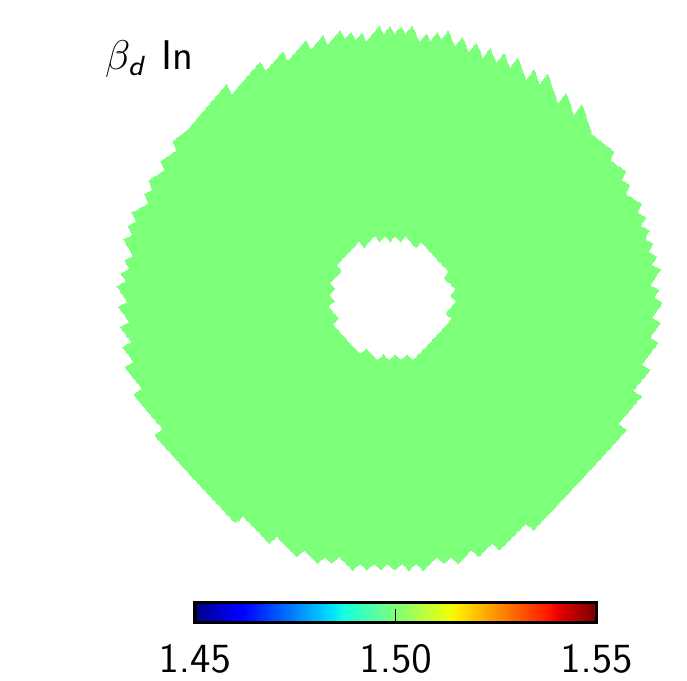}}
\subfigure{\includegraphics[width=0.208\linewidth,clip=true,trim=0.3in 0in 0.05in 0.1in]{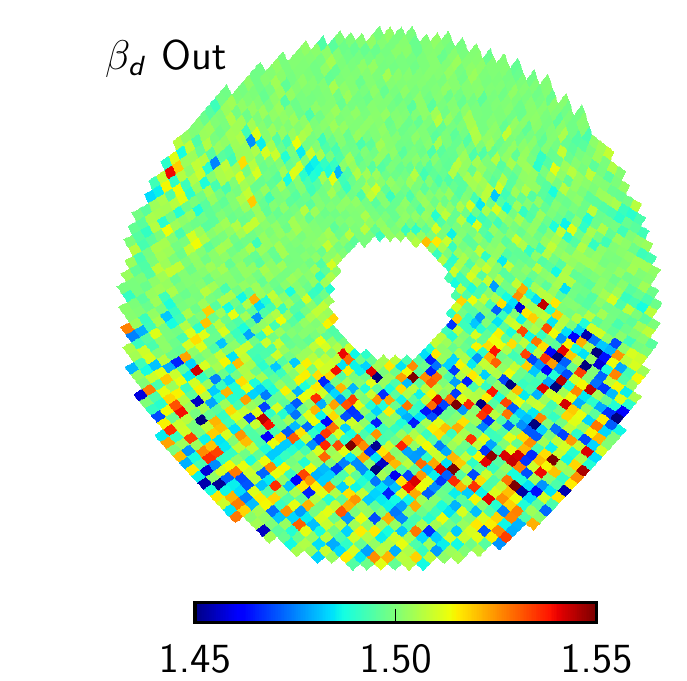}}
\subfigure{\includegraphics[width=0.208\linewidth,clip=true,trim=0.3in 0in 0.05in 0.1in]{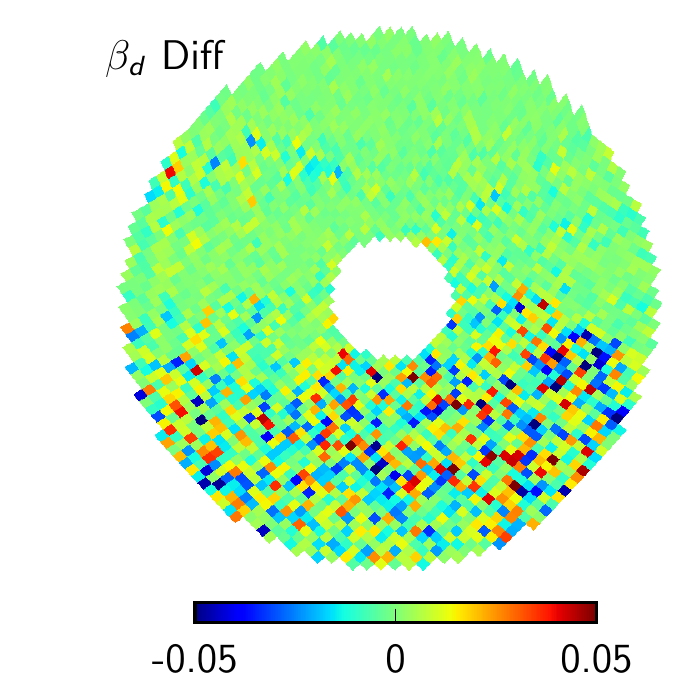}}
\subfigure{\includegraphics[width=0.208\linewidth,clip=true,trim=0.3in 0in 0.05in 0.1in]{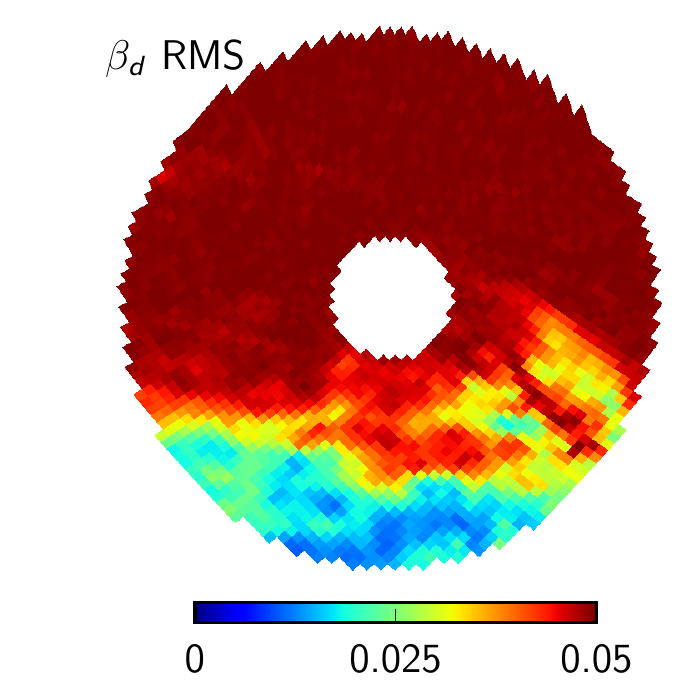}}
\vspace{-0.05in}
\caption{Comparison of input maps (first column), reconstructed
  maps (second column), residual (output minus input) maps
  (third column), and posterior RMS maps (fourth column)
  for the baseline SN10 experimental setup. From top to
  bottom, rows show: (1) Stokes $Q$ CMB in thermodynamic units, (2) Stokes $Q$ synchrotron
  amplitude at 23~GHz in Rayleigh-Jeans (RJ) units, (3) Stokes $Q$ thermal dust
  amplitude at 353~GHz in RJ units, (4) synchrotron spectral index, and
  5) thermal dust spectral index. }
\label{fig:diffmaps}
\end{figure*}

\begin{figure}[t]
\begin{center}
\mbox{\includegraphics[width=\linewidth,clip=]{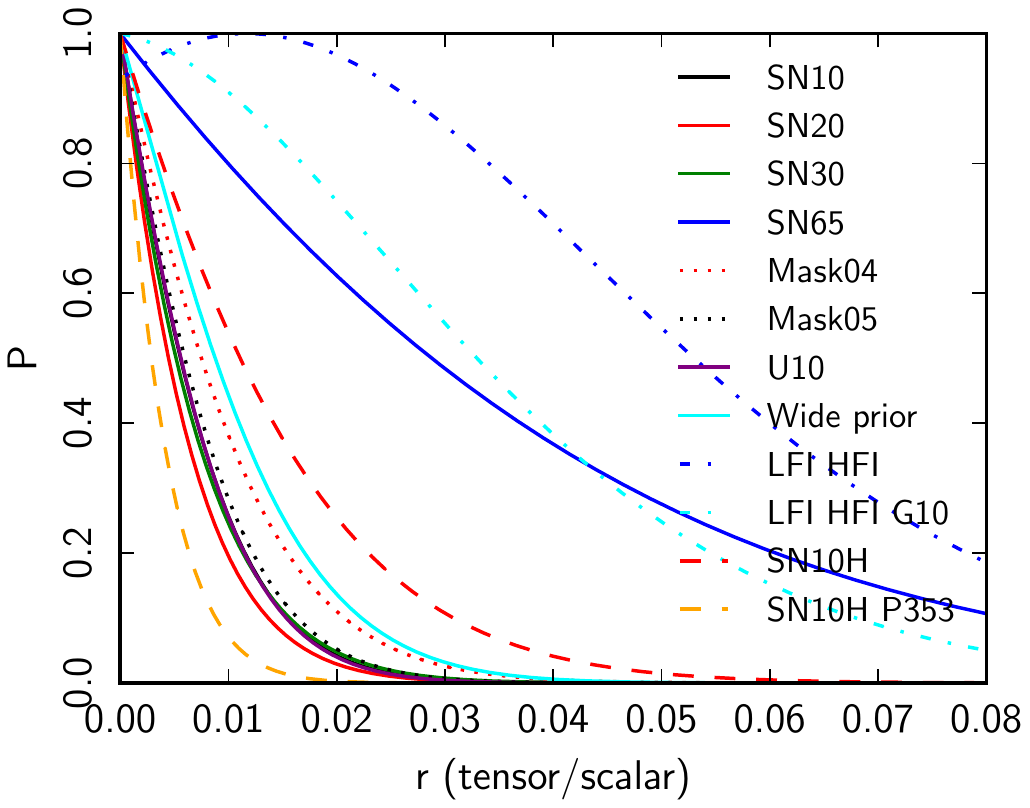}}
\end{center}
\caption{Marginal tensor-to-scalar ratio posterior distributions for various
  experimental setups, averaged over 20 simulations. All setups, except
 SN10H and SN10H G10, contain the \Planck\ HFI channels, and all except LFI HFI and LFI HFI G10, contains the 5
 lowest \name\ channels.}
\label{fig:r_curves_rms}
\end{figure}

\section{Results}
\label{sec:results}

With the above simulation and analysis pipeline, we are now ready to
explore the effect of various experimental setups on our final
constraints on the tensor-to-scalar ratio. 

\subsection{Baseline configuration}
\label{sec:baseline}

Our first simulation configuration represents the baseline against which 
all other configurations are compared. This setup adopts the SN10
observational strategy, resulting in a sky fraction of
$f_{\mathrm{sky}}=0.06$, and adopting signal-to-noise weighting for
the \name\ frequency bands, see Table~\ref{tab:instrument}. The foreground model includes synchrotron
and thermal dust emission both in the simulation and in the analysis,
with the product of uniform ($-4.5 < \beta_{\mathrm{s}} < -1.5$ and
$1.0 < \beta_{\mathrm{d}} < 2.0$) and Gaussian 
($\beta_{\mathrm{s}}\sim\mathcal{N}(-3.1,\sigma_{\beta_{\mathrm{s}}}=0.1)$ and 
$\beta_{\mathrm{d}}\sim\mathcal{N}(1.50,\sigma_{\beta_{\mathrm{d}}}=0.05)$ )
priors. The tensor-to-scalar ratio
is fixed to $r=0$, while all other cosmological parameters are set to
the best-fit $\Lambda$CDM model described in Sect.
\ref{sec:simulations}.

In Fig.~\ref{fig:diffmaps} we compare the true input maps (left column) with
the reconstructed posterior mean maps (second column) for each free
parameter for one simulation with this particular setup. We only show
Stokes $Q$ in this plot, since $U$ looks qualitatively very
similar. The third column shows their difference, while the fourth
column shows the posterior RMS map as estimated by
the Gibbs sampler; ideally, the maps in the third column should be
consistent with Gaussian random noise with a standard deviation given
by the fourth column. At least visually, this expectation appears to
be met quite well, with no obvious signs of consistent correlated
structures in the residual maps. Of course, since the fitted model
matches the true signal, this is not surprising, but it still serves
as an important self-consistency check.

The structures of the posterior RMS maps also agree with our
expectations. For instance, the uncertainty in the component amplitude
parameters (CMB, synchrotron, and thermal dust) is larger where the
foregrounds are bright, whereas the uncertainty in the spectral index
parameters is smaller in the same region. For the CMB, this spatially
varying standard deviation naturally down-weighs regions with bright
foregrounds in the tensor-to-scalar ratio likelihood fit, serving
effectively as a mild form of masking. Along the edges of the amplitude maps, the uncertainties are generally smaller because of the relatively higher number of observations per pixel due to the scanning pattern, this can be seen in the CMB and synchrotron RMS maps. For the thermal dust RMS map, the sharp features trace the \Planck\ scanning strategy.

The black solid line in Fig.~\ref{fig:r_curves_rms} shows the
resulting marginal tensor-to-scalar ratio posterior distribution,
averaged over 20 simulations. This distribution peaks at $r=0$, in
agreement with the input value, and has an upper 95\% confidence limit
of $r=0.017$. This value thus represents a baseline comparison point
against which we can compare other experimental setups.

Before doing that, however, it is interesting to understand at what
level foreground marginalization degrades the overall results,
compared to a CMB-only setup with the same sensitivity. To answer this
question, we compare in Fig.~\ref{fig:r_cmb} the upper 95\% confidence
limits averaged over 20 simulations with the corresponding $68\%$ region error bars, as derived with four simple sky model variations, namely (1) CMB
only; (2) CMB and thermal dust; (3) CMB and synchrotron; and (4) CMB,
synchrotron, and thermal dust (all in addition to noise). The latter point thus corresponds
directly to the baseline model. Overall, we see that all four models
result in very similar constraints, with a very slight posterior
degradation of about 10\% due to synchrotron emission marginalization,
but no measurable degradation due to thermal dust
marginalization. The fact that the mean point is slightly lower in
the CMB+dust model is due to the limited number of simulations, and is
not statistically significant. This shows that the frequency band
selection in the baseline configuration results in a robust component
separation efficiency, achieving nearly optimal post-component
separation results.

As a bias test and sanity check on the baseline configuration, we have
also analyzed a similar set of simulations, but this time with
adopting a nonzero value of the tensor-to-scalar ratio of $r=0.05$
and fixed spectral indices, $\sigma_{\mathrm{d}} = \sigma_{\mathrm{s}} = 0$; 
the latter choice is introduced because
the mathematical problem then has an algebraically closed solution,
which is useful for validation. The results from this calculation are
shown in Fig.~\ref{fig:r_050}; as expected, the recovered posterior
distribution is consistent with the input value. 
We note that the width of this distribution is larger for $r=0.05$ 
than for $r=0$, as  expected due to cosmic variance.

\subsection{Configuration variations}

We are now ready to explore the larger experimental parameter space,
introducing variations around the baseline model SN10. In each
case, we only focus on the resulting tensor-to-scalar ratio posterior
distributions, and in particular on the upper 95\% confidence limits,
and not show individual sky maps for each; these look qualitatively
similar to those shown in Fig.~\ref{fig:diffmaps}. In 
Table~\ref{tab:configuration} all the different configuration
variations are listed with check marks at the relevant \name\ and 
\Planck\ Low Frequency Instrument (LFI) and High Frequency Instrument (HFI) frequency channels,
together with the resulting upper 95\% confidence limits on $r$.

\begin{figure}[t]
\begin{center}
\mbox{\includegraphics[width=\linewidth,clip=]{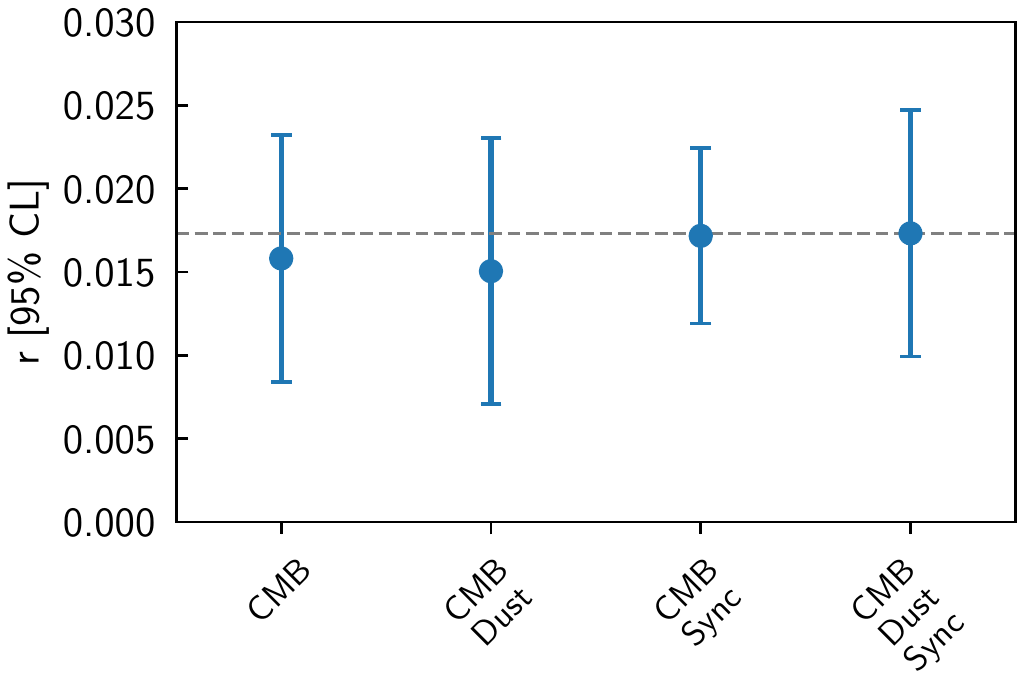}}
\end{center}
\caption{Comparison of the upper $95\%$ confidence limits of the
  tensor-to-scalar ratio posteriors for (1) CMB only; (2) CMB and
  thermal dust; (3) CMB and synchrotron; and (4) CMB, synchrotron, and
  thermal dust (in addition to noise), all evaluated for the baseline experiment
  configuration. Each point represents the mean of the 95\%
  confidence limits evaluated from 20 simulations, and the error bar
  indicates the 68\% region among the same simulations. The horizontal
  gray dashed line indicates the value for Case 4, representing the
  full model, SN10. The true input value is $r=0$.}
\label{fig:r_cmb}
\end{figure}

\begin{figure}[t]
\begin{center}
\mbox{\includegraphics[width=\linewidth,clip=]{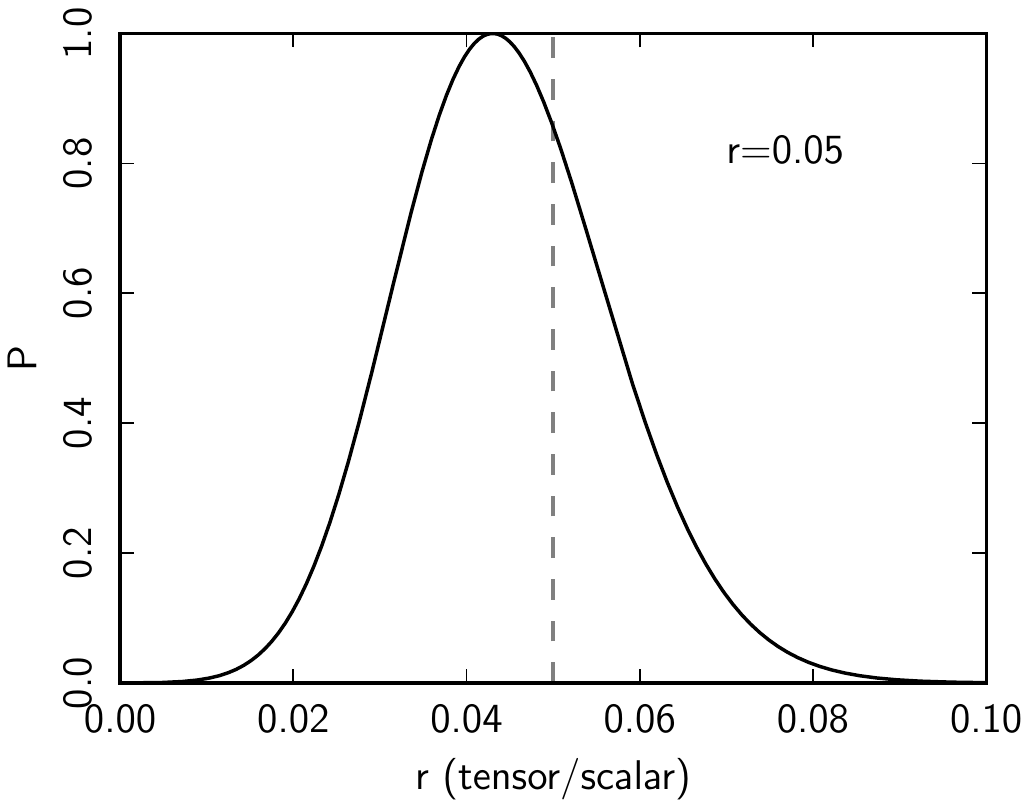}}
\end{center}
\caption{Marginal tensor-to-scalar ratio posterior distribution for the case
  where $r=0.05$ and fixed spectral index priors $\sigma_{\mathrm{d}} = \sigma_{\mathrm{s}} = 0$, 
  averaged over 20 simulations. The vertical gray dashed line is the true value,
  $r=0.05$.}
\label{fig:r_050}
\end{figure}

\begin{table*}[t]                          
\begingroup        
\newdimen\tblskip \tblskip=2pt
\caption{Upper 95\% confidence limits (CL) of r and included frequency channels for each experimental configuration. The limits are the mean of 20 simulations, and the error is the one sigma ($68\%$) region. The input values for all configurations except the bottom one are $r=0$, and the recovered values are all consistent with input. 
\label{tab:configuration} }
\nointerlineskip                                                     
\vskip -3mm
\footnotesize     
\setbox\tablebox=\vbox{        
\newdimen\digitwidth       
\setbox0=\hbox{\rm 0}
\digitwidth=\wd0
\catcode`*=\active
\def*{\kern\digitwidth}
\newdimen\signwidth
\setbox0=\hbox{+}
\signwidth=\wd0
\catcode`!=\active
\def!{\kern\signwidth}
\newdimen\decimalwidth
\setbox0=\hbox{.}
\decimalwidth=\wd0
\catcode`@=\active
\def@{\kern\signwidth}
%
\halign{#\hfil\tabskip=1.0em&
    \hfil#\hfil\tabskip=1em&
    \hfil#\hfil\tabskip=0.3em&
    \hfil#\hfil\tabskip=0.3em&
    \hfil#\hfil\tabskip=0.3em&
    \hfil#\hfil\tabskip=0.3em&
    \hfil#\hfil\tabskip=0.3em&
    \hfil#\hfil\tabskip=0.3em&
    \hfil#\hfil\tabskip=2.em&
    \hfil#\hfil\tabskip=0.3em&
    \hfil#\hfil\tabskip=0.3em&
    \hfil#\hfil\tabskip=2em&
    \hfil#\hfil\tabskip=0.3em&
    \hfil#\hfil\tabskip=0.3em&
    \hfil#\hfil\tabskip=0.3em&
    \hfil#\hfil\tabskip=0.3pt\cr
\noalign{\doubleline}
&&\multispan{7}{\hfil\bf\name\hfil}& \multispan{3}{\hfil\bf LFI\hfil}& \multispan{4}{\hfil\bf HFI\hfil}\cr
\noalign{\vskip 3pt\hrule\vskip 8pt}
Configuration& $r$ [95\% CL] & 10 & 15 & 20 & 30 & 44 & 90 & 143 & 30 & 44 & 70 & 100 & 143 &217 & 353 \cr
\noalign{\vskip 3pt\hrule\vskip 8pt}
SN10 &$0.017\pm0.007$ &\checkmark&\checkmark&\checkmark&\checkmark&\checkmark&&&&&&\checkmark&\checkmark&\checkmark&\checkmark \cr
SN20 &$0.016\pm0.006$ &\checkmark&\checkmark&\checkmark&\checkmark&\checkmark&&&&&&\checkmark&\checkmark&\checkmark&\checkmark \cr
SN30 &$0.019\pm0.004$ &\checkmark&\checkmark&\checkmark&\checkmark&\checkmark&&&&&&\checkmark&\checkmark&\checkmark&\checkmark \cr
SN65 &$0.09\pm0.02$ &\checkmark&\checkmark&\checkmark&\checkmark&\checkmark&&&&&&\checkmark&\checkmark&\checkmark&\checkmark \cr
Mask04 &$0.023\pm0.009$ &\checkmark&\checkmark&\checkmark&\checkmark&\checkmark&&&&&&\checkmark&\checkmark&\checkmark&\checkmark \cr
Mask05 &$0.018\pm0.008$ &\checkmark&\checkmark&\checkmark&\checkmark&\checkmark&&&&&&\checkmark&\checkmark&\checkmark&\checkmark \cr
U10 &$0.017\pm0.007$ &\checkmark&\checkmark&\checkmark&\checkmark&\checkmark&&&&&&\checkmark&\checkmark&\checkmark&\checkmark \cr
Wide prior &$0.023\pm0.009$ &\checkmark&\checkmark&\checkmark&\checkmark&\checkmark&&&&&&\checkmark&\checkmark&\checkmark&\checkmark \cr
LFI HFI &$0.08\pm0.03$ &&&&&&&&\checkmark&\checkmark&\checkmark&\checkmark&\checkmark&\checkmark&\checkmark \cr
LFI HFI G10 &$0.07\pm0.03$ &\checkmark&&&&&&&\checkmark&\checkmark&\checkmark&\checkmark&\checkmark&\checkmark&\checkmark \cr
SN10H &$0.034\pm0.009$ &\checkmark&\checkmark&\checkmark&\checkmark&\checkmark&\checkmark&\checkmark&&&&&&& \cr
SN10H P353 &$0.011\pm0.002$ &\checkmark&\checkmark&\checkmark&\checkmark&\checkmark&\checkmark&\checkmark&&&&&&&\checkmark \cr
Spinning dust, Mask05 &$0.026\pm0.010$ &\checkmark&\checkmark&\checkmark&\checkmark&\checkmark&&&&&&\checkmark&\checkmark&\checkmark&\checkmark \cr
r=0.05 & $\cdots$ &\checkmark&\checkmark&\checkmark&\checkmark&\checkmark&&&&&&\checkmark&\checkmark&\checkmark&\checkmark \cr
\noalign{\vskip 2pt\hrule\vskip 2pt}
}}
\endPlancktablewide
\endgroup
\end{table*}

\begin{figure*}[t]
\begin{center}
\mbox{\includegraphics[width=\linewidth,clip=]{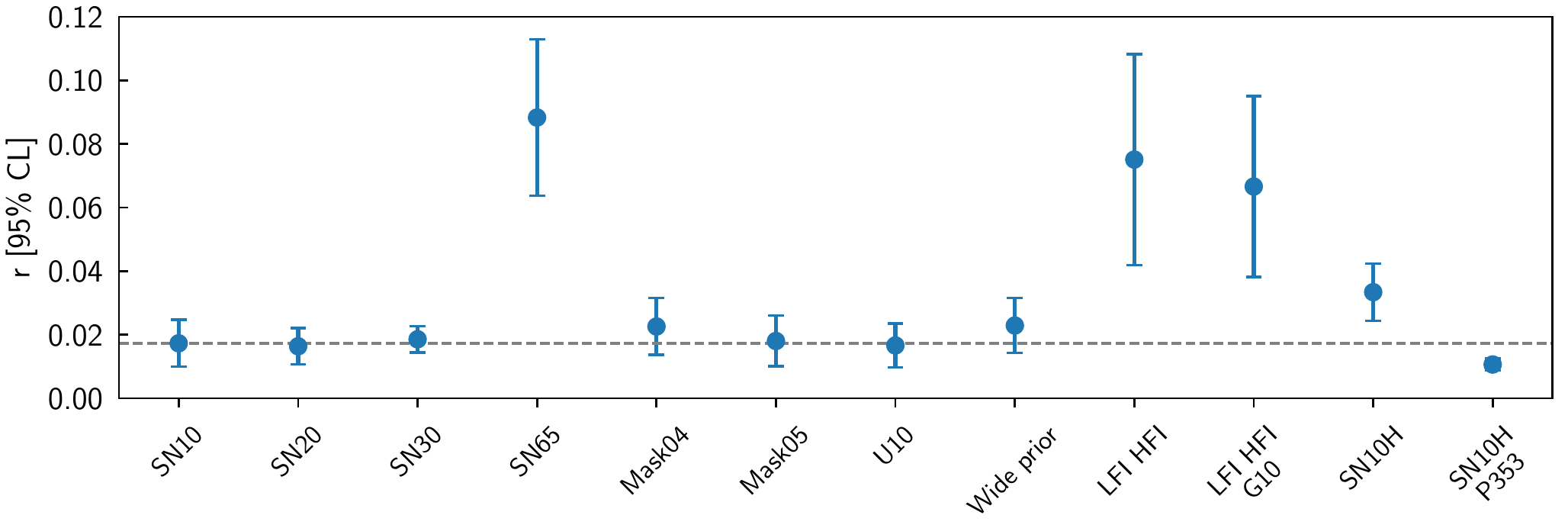}}
\end{center}
\caption{Comparison of the upper $95\%$ confidence limits of the
  tensor-to-scalar ratio $r$ for various experimental setups. Each
  point represents the mean of the 95\% confidence limits evaluated
  from 20 simulations, and the error bar indicates the 68\% region
  among the same simulations. The horizontal gray dashed line is the
  value of the baseline configuration, SN10. The true input value for
  all these configurations is $r=0$.}
\label{fig:r_curves_whiskers}
\end{figure*}

\paragraph{\emph{Sky fraction}} --- In our first set of variations, we simply
change the effective sky fraction, adopting opening angles of
$\alpha=20^{\circ}$, $\alpha=30^{\circ}$, and $\alpha=65^{\circ}$,
corresponding to sky fractions of 0.11, 0.16 and 0.44,
respectively. The cases are labeled SN20, SN30, and SN65.
The posterior distributions obtained from these are
shown as red, green and blue solid lines in
Fig.~\ref{fig:r_curves_rms}, and the resulting upper 95\% confidence
limits are plotted with one sigma error bars in
Fig.~\ref{fig:r_curves_whiskers}. Overall, we see that a larger sky fraction 
generally leads to weaker constraints on the tensor-to-scalar ratio, 
although for sky fractions
up to 0.16 the effect is very marginal. For the largest sky fraction
of 0.44, however, the effect is large, effectively reducing the
detection power by a factor of four. The main reason for this dramatic
increase is simply that for the largest opening angle, a relatively
large fraction of the observation time is spent within the Galactic
Plane, where the CMB constraints are relatively poor.

\paragraph{\emph{Masking}} --- To quantify the effect of masking on
the smaller fields, we next consider two cases in which
high-foreground pixels are simply removed from the analysis by an
explicit mask. Two different masks are considered. Mask04 is defined by
thresholding the \Planck\ synchrotron polarization, resulting in a sky 
fraction of $f_\mathrm{sky}=0.73$ for a full sky map, 
and $f_\mathrm{sky}=0.04$ for our $10^{\circ}$ field.
The other mask, Mask05, is defined by thresholding the CMB posterior 
RMS map at $0.4\,\mu\textrm{K}$ (and thus only removing pixels in the Stokes 
$Q$ map), resulting in a sky fraction of $f_\mathrm{sky}=0.05$ averaged 
over Stokes $Q$ and $U$. The resulting posteriors are shown as dotted lines
in Fig.~\ref{fig:r_curves_rms}, and the upper 95\% confidence limits
plotted in Fig.~\ref{fig:r_curves_whiskers}. As expected, the
constraints degrade proportionally to the removed sky fraction.

\paragraph{\emph{Noise weighting}} --- Next, we consider a different relative
weighting between frequency channels. As described above, the baseline
configuration adopts total signal-to-noise ratio weights for each
frequency channel (i.e., number of detectors per band), which is
qualitatively similar to \Planck. In the next variation, we instead
consider uniform weighting between frequencies, U10, as
measured in units of telescope years. In terms of implementation
strategy, this is easier to achieve, since each telescope can then be
tuned to a specific frequency, and does not require multi-frequency
optical elements. The resulting tensor-to-scalar ratio posterior
distribution and upper 95\% confidence limit are summarized in
Figs.~\ref{fig:r_curves_rms} and \ref{fig:r_curves_whiskers}. Overall,
this configuration performs very similar to the baseline
configuration, suggesting that the specific choice of detector weights
is not critical for the final results. This, of course, is not
surprising, considering the stability of the CMB solution with respect
to foregrounds that was demonstrated in Fig.~\ref{fig:r_cmb}.

\paragraph{\emph{Foreground priors}} --- As a further test of foreground
stability, we next consider the case in which the Gaussian priors on
the spectral parameters are loosened (tripled), from 
$\sigma_{\beta_\textrm{s}}=0.1$ and $\sigma_{\beta_\textrm{d}} = 0.05$ to 
$\sigma_{\beta_\textrm{s}}=0.3$ and $\sigma_{\beta_\textrm{d}}=0.15$, respectively. 
The resulting posterior distribution is plotted as a cyan solid curve in
Fig.~\ref{fig:r_curves_rms} (Wide prior), and the upper 95\% confidence limit on
the tensor-to-scalar ratio degrades by about 30\%, from $r<0.020$ to
$r<0.025$.

\begin{figure}[t]
\begin{center}
  \mbox{\includegraphics[width=0.49\linewidth,clip=]{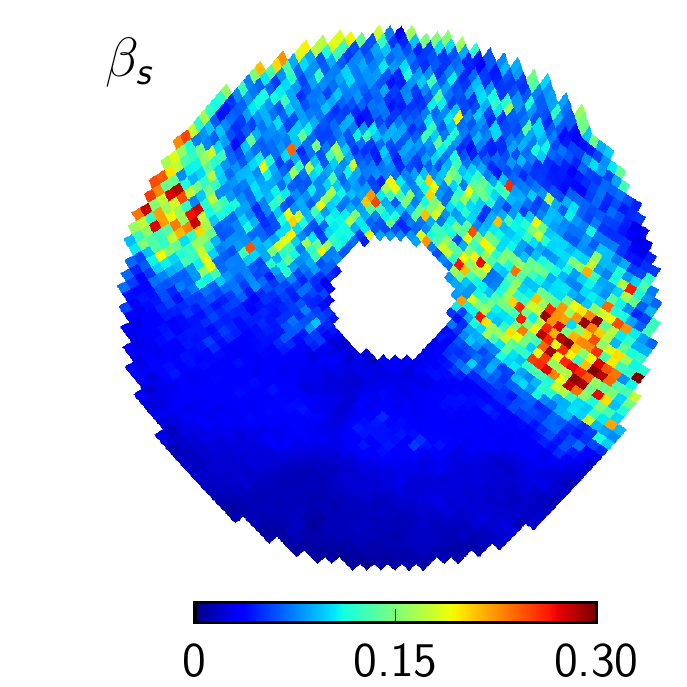}}
  \mbox{\includegraphics[width=0.49\linewidth,clip=]{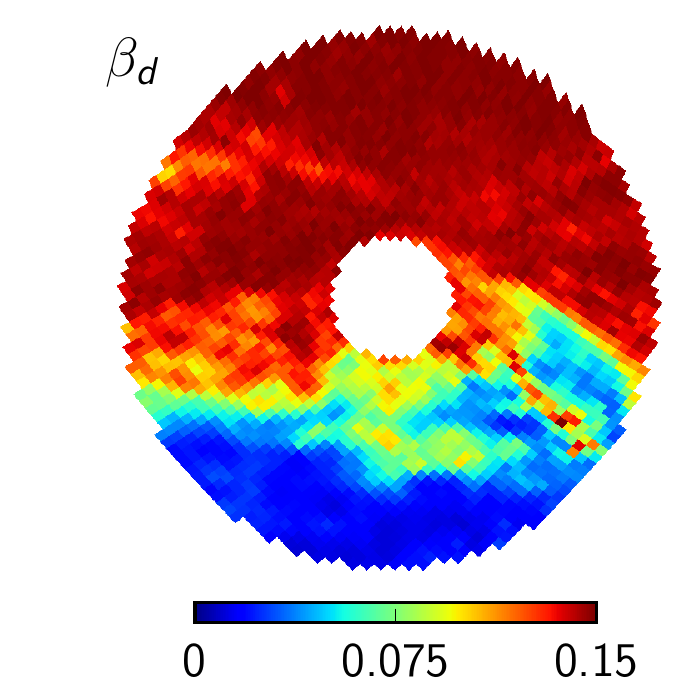}}
  \mbox{\includegraphics[width=0.49\linewidth,clip=]{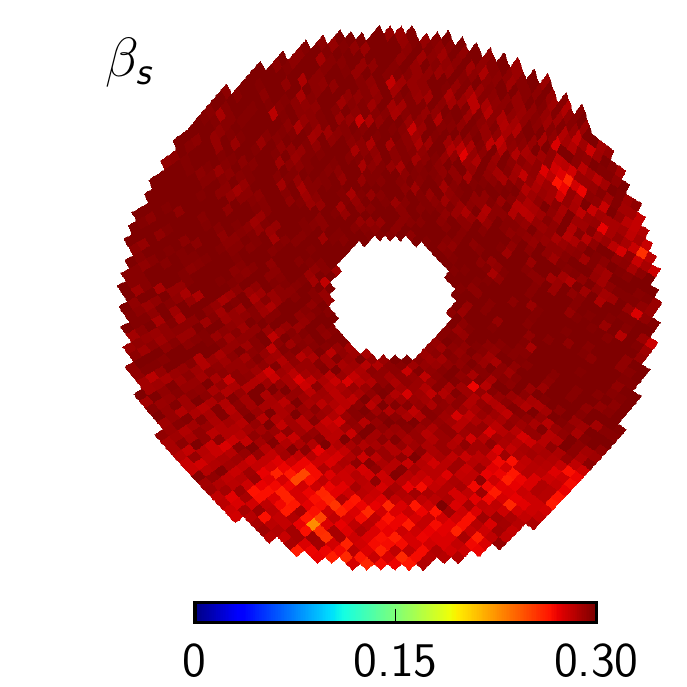}}
  \mbox{\includegraphics[width=0.49\linewidth,clip=]{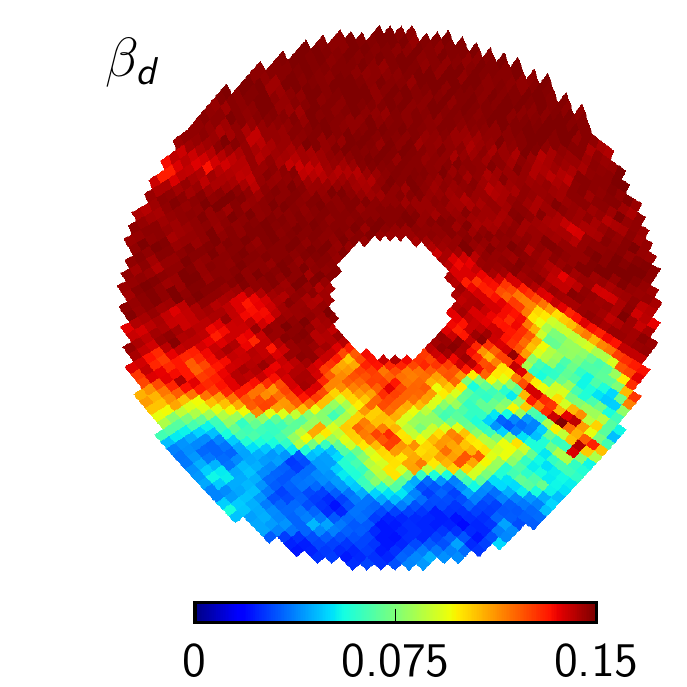}}
\end{center}
\caption{Comparison of RMS maps of the synchrotron (left column) and dust (right column) 
spectral indices for two different configurations.
    Top: Posterior RMS maps for the Wide prior setup 
  with the 5 low \name\ frequencies plus \Planck\ HFI.
  Bottom: Posterior RMS maps for the \Planck\ LFI HFI setup 
  sampled with a wide prior.
  }
\label{fig:synch_dust}
\end{figure}

\paragraph{\emph{Comparison with other experiments}} --- We
now consider combinations of different data sets and frequencies. In
the first case (LFI HFI), we simply replace all \name\ frequencies with the
three currently available \Planck\ LFI frequencies at 30, 44 and
70~GHz. As seen in Fig.~\ref{fig:r_curves_whiskers}, this results in a
mean upper 95\% confidence limit on the tensor-to-scalar ratio of
$r<0.08$. Adding only the 10~GHz \name\ channel to \Planck\ LFI+HFI
combination (LFI HFI G10) improves this limit to $r<0.07$. In the third case, we
replace the \Planck\ HFI frequencies between 100 and 353~GHz with
internal \name\ bands at 90 and 143~GHz, SN10H; this results in
a limit of $r<0.035$. The fourth and last case (SN10H P353) is one where we add
only the \Planck\ 353~GHz channel to the full-range
\name\ configuration; this results in the tightest limit of all cases
considered in this paper, resulting in a 95\% confidence limit of
$r<0.012$. Overall, it is clear that internal high-frequency channels
are not able to replace the \Planck\ HFI data, and in particular the
353~GHz channel, in terms of sheer sensitivity, but they would be
extremely valuable additions both in terms of final sensitivity and
overall robustness with respect to instrumental systematics.

In Fig.~\ref{fig:synch_dust} we show the posterior RMS maps for synchrotron
and thermal dust spectral indices for the Wide prior setup containing five low \name\ channels plus HFI (top)
and the \Planck\ only setup, (bottom).
Both setups are sampled using wide priors, $\sigma_{\beta_\textrm{s}}=0.3$ and $\sigma_{\beta_\textrm{d}}=0.15$. 
Here we clearly see that the \Planck\ setup is totally
prior dominated with regard to the synchrotron spectral index, while the setup
containing low \name\ frequencies instead of LFI, is able to get good constraints on the spectral index. For the case of thermal dust spectral index, both setups are prior dominated at the highest latitudes where the signal is low. In fact the setup containing the low \name\ frequencies is even a little bit less prior dominated than the \Planck\ only setup.

\begin{figure}[t]
\begin{center}
  \mbox{\includegraphics[width=\linewidth,clip=]{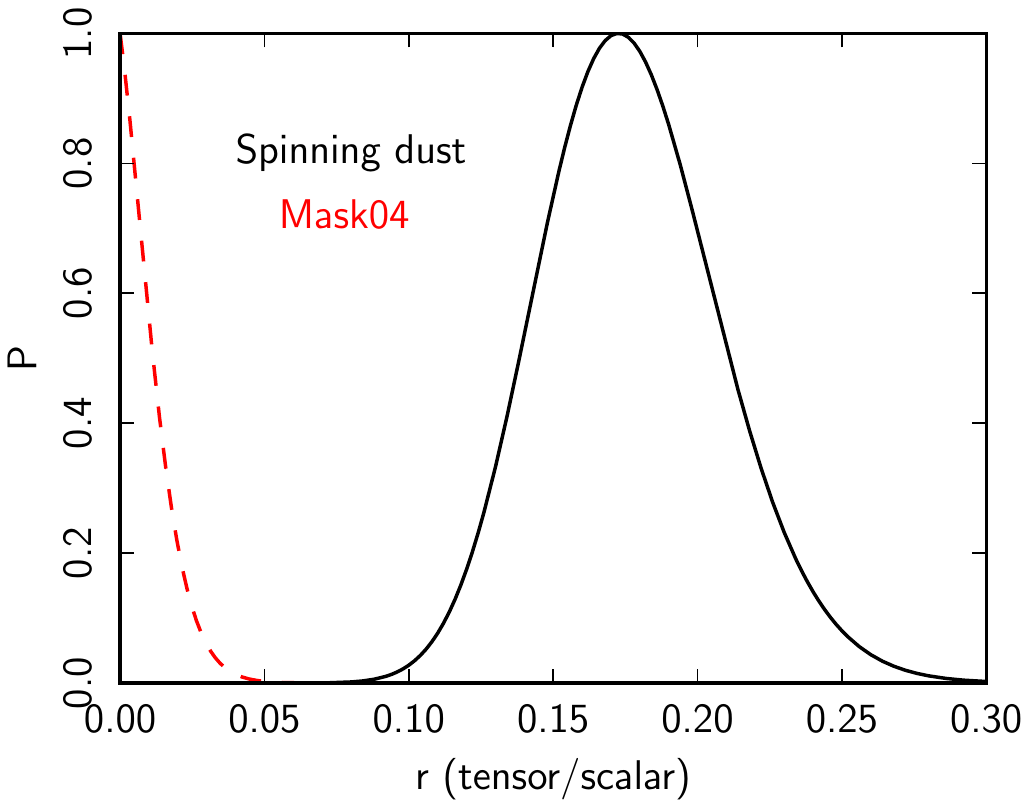}}
  \mbox{\includegraphics[width=0.7\linewidth,clip=]{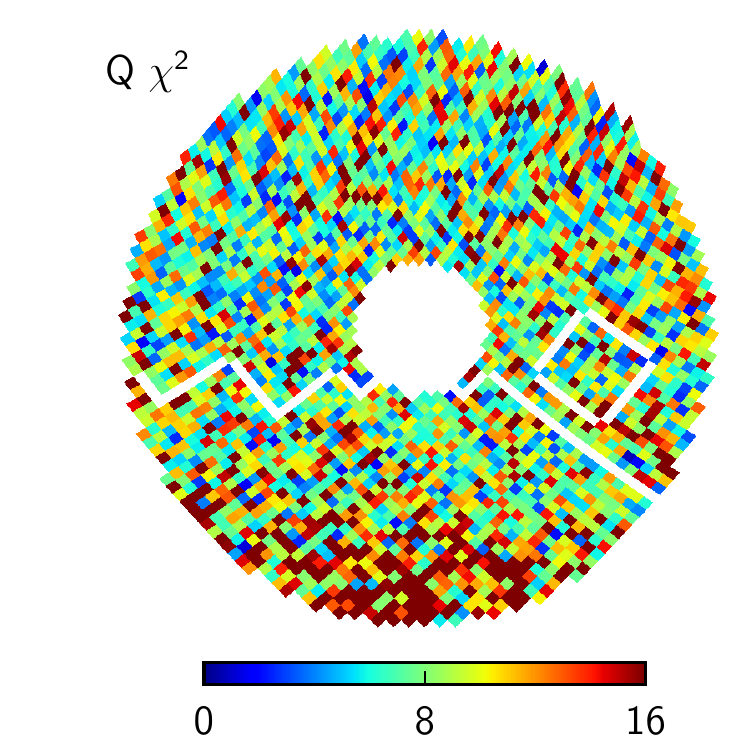}}
\end{center}
\caption{Results for the modeling errors case.
    Top: Marginal tensor-to-scalar posterior distribution
  for a model mismatch of adding a spinning dust component in the
  simulation. The dashed red line shows the posterior after the mask, Mask04,
  is applied. Bottom: Corresponding Stokes $Q$ $\chi^2$ map for
  one arbitrary sample. The white line indicates the masked-out region.}
\label{fig:spindust}
\end{figure}

\paragraph{\emph{Modeling errors}} --- Finally, we consider a case with
an explicit mismatch between the data and the fitted model, which is
generated by adding a spinning dust component with a polarization
fraction of 1\% to the input simulation. The resulting
tensor-to-scalar ratio posterior is shown in the top panel of
Fig.~\ref{fig:spindust}. In this case, the result is formally a
spurious foreground-induced detection of a tensor-to-scalar ratio of
$r=0.16\pm0.04$. The bottom panel of Fig.~\ref{fig:spindust}
shows the goodness of fit, the $\chi^2$ map (Stokes $Q$) resulting from this analysis; the
corresponding reduced $\chi^2$ is $1.22\pm0.02$, indicating the
presence of a strong model mismatch as opposed to being unity as 
in all the previous cases. Thus, this spurious detection
would be rejected internally by the component separation process.
When we apply a mask on the foreground dominated part, Mask04, 
indicated by the white line on the $\chi^2$ map, 
the detection is replaced by a value that fits a vanishing $r$, 
with a 95\% confidence limit of $r<0.03$.
The red dashed line in Fig.~\ref{fig:spindust} shows the corresponding 
posterior curve.

\section{Conclusions}
\label{sec:conclusions}

In this paper, we have performed a preliminary sensitivity analysis
for \name, a hypothetical new ground-based CMB $B$-mode polarization
experiment. This experiment is designed to observe the Northern
Galactic Hemisphere from the Summit site on Greenland; at an altitude
of 3216 m above sea level and a latitude of $72^{\circ}$N. This
site exhibits excellent observing conditions for future CMB
polarization experiments, both in terms of atmospheric transmission
and observing strategies. \name\ would then primarily observe at
low-frequencies, between 10 and 44 GHz, and combine the resulting data
with observations from the \Planck\ HFI instrument and/or internal
observations at 90 and 143~GHz. The proposed \name\ detector technology and
unique observing site makes the experiment complementary to existing
efforts, most of which focus on the Southern Galactic Hemisphere.

The primary goal of the current paper is to quantify the sensitivity
of \name\ in terms of constraints on the tensor-to-scalar ratio after
component separation. We have done this by analyzing simplified and
controlled simulations within a well-established Bayesian component
separation framework called Commander.
We find that the constraints on synchrotron emission improve when adding the low frequency \name\ data to the \Planck\ data. We expect that this trend will become even more important when also accounting for spatial variations in the spectral index and curvature; however, exploring this is beyond the scope of this paper.

To summarize, our main conclusions are the following:
\begin{enumerate}
\item When combined with \Planck\ HFI observations, \name\ would constrain the tensor-to-scalar ratio to $r<0.02$ at 95\% confidence
  limit. This is competitive with existing constraints. For instance, the current best
  constraint from BICEP2+Keck+\Planck\ is $r<0.032$ \citep{tristram:2022}.
\item The value of $r<0.02$ is robust with
  respect to specific details in the experimental configuration;
  neither sky fraction (up to a certain limit), detector weighting,
  nor foreground priors affect this value significantly. Indeed, the
  frequency layout proposed for \name\ provides a foreground stability
  that yields constraints on the tensor-to-scalar ratio that exceeds
  the CMB-only prediction by only about 10\%.
\item For cross-validation purposes, the \Planck\ HFI observations could be replaced with internal
  observations at 90 and 143\,GHz. This carries a cost of about 100\% in
  constraining power, increasing the tensor-to-scalar ratio limit to
  $r<0.035$. However, having two fully independent data sets available
  for thermal dust reconstruction would also greatly improve our
  ability to reject a potential false detection due to instrumental
  systematic effects.
\end{enumerate}

\begin{acknowledgements}
Some of the results in this paper have been derived using the \emph{healpy} 
\citep{zonca:2019} and HEALPix \citep{gorski:2005} packages. 
We are grateful to Villum Fond for support of the Deep Space project. 
This work has received funding from the European
  Union’s Horizon 2020 research and innovation programme under grant
  agreement numbers 819478 (ERC; \textsc{Cosmoglobe}) and 772253 (ERC; \textsc{bits2cosmology}). This work is supported by the Research Council of Norway under grant agreement number 230947.
This work is also supported in part by the National Key R\&D Program of China
(2021YFC2203100, 2021YFC2203104) and the Anhui project Z010118169. 
SvH acknowledges funding from the Beecroft Trust.
AK, PM and PL were supported by the Niels Bohr Institute and the Villum Foundation. We would like to thank the US NSF and the staff at Summit Station for excellent support during our observing campaign.
\end{acknowledgements}

\end{document}